%% file: main.tex
\documentclass[
  aps,
  pra,
  twocolumn,
  superscriptaddress,
  10pt,
  longbibliography
]{revtex4-2}

\usepackage{packages/ignore}
\usepackage{packages/main}
\usepackage{packages/eq}
\graphicspath{{figs/}}
\DeclareGraphicsExtensions{.pdf,.png,.jpg,.jpeg}

\allowdisplaybreaks[4]
\begin{document}

\title{Floquet Driving of Enzymatic Reactions: Counting Statistics and Long-Time Currents}

\author{Yuki Watanabe}
\affiliation{The Institute for Solid State Physics, The University of Tokyo, 5-1-5 Kashiwanoha, Kashiwa, Chiba 277-8581, Japan}

\author{Yuki Ishiguro}
\affiliation{Faculty of Engineering, Tokyo Polytechnic University, 5-45-1 Iiyama-minami, Atsugi, Kanagawa 243-0297, Japan}
\affiliation{The Institute for Solid State Physics, The University of Tokyo, 5-1-5 Kashiwanoha, Kashiwa, Chiba 277-8581, Japan}
\author{Takashi Oka}
\affiliation{The Institute for Solid State Physics, The University of Tokyo, 5-1-5 Kashiwanoha, Kashiwa, Chiba 277-8581, Japan}

\date{\today}
\input{sections/0-abstract}

\maketitle
\input{sections/1-introduction}
\input{sections/2-model}
\input{sections/3-floquet-theory}
\input{sections/4-application-for-discrete-driving-protocols}
\input{sections/5-conclusion}

\section*{Acknowledgments}
We would like to thank Ryusuke Hamazaki, Kohei Yoshimura, Akihiro Ozawa, Hung-Hsuan Teh, Yuuki Sugiyama, Takahiro Sagawa, Kumiko Hayashi, Keiichi Inoue, and Takashi Nagata for fruitful discussions.
Y.W. is supported by MERIT-WINGS of the University of Tokyo.
This work is supported by JSPS KAKENHI Grant No.~JP23H04865, JP23K25837 and JP24K16976.
\bibliography{discrete-floquet-theory}

\input{sections/10-appendix}

\end{document}

%% file: sections/0-abstract.tex
\begin{abstract}
Technologies for artificially controlling chemical reaction systems, such as optogenetics, are rapidly advancing, making it increasingly important to understand reaction dynamics under time-dependent control. When the modulation of reaction rates is periodic in time, the Floquet formalism provides a systematic framework. We develop a Floquet theory for classical stochastic processes that enables the calculation of the current and its counting statistics under such periodic modulation. In particular, we formulate the theory in terms of a counting field and derive general expressions for the first cumulant and the corresponding current. The current is expressed using the effective Floquet generator and the kicked state, and we further obtain general asymptotic expressions for the current in both the high- and low-frequency regimes. As a concrete example to test our analytical expressions, we then apply the results to discrete Floquet driving---a non-perturbative, stepwise protocol. The setup is motivated by a biochemical system known as cyclic adenosine monophosphate (cAMP) production, which is an enzymatic reaction activated and inhibited by G-proteins. This is formulated as a discretely driven Michaelis--Menten-type reaction model, in which the catalytic activity is switched on and off abruptly in time, and we obtain analytical expressions and numerical results showing how periodic switching of reaction rates generates a long-time product current. In particular, in the high-frequency limit, we show that the effect of the periodic driving can be interpreted through an effective modification of the chemical reaction rates. These results provide a basis for Floquet analysis of periodically driven chemical reactions.
\end{abstract}

%% file: sections/1-introduction.tex

\section{Introduction} \label{sec:introduction}
Classical stochastic processes describe systems evolving randomly over time and provide a framework to understand the dynamics driven by probabilistic reaction events and transitions~\cite{Van-Kampen2007-hz,Gardiner2009-nr,Risken1996-wl,Helbing2001-tu}. Various examples have been studied, ranging from traffic flow~\cite{Derrida1998-yw,Schadschneider2010-rh} to fluctuation statistics in non-equilibrium systems~\cite{Seifert2012-jn}.
Stochastic systems in {\it time-varying} environments are of growing interest~\cite{Jung1993-rs,Sato2025-gh,Ohkubo2008-xl,Ohkubo2008-za,Sinitsyn2007-qo,Sinitsyn2007-uj,Sinitsyn2009-jy,Rahav2008-kx}.
In chemical reaction systems, in particular, reaction rates are often subject to temporal modulation, either through external control or through coupling to internal degrees of freedom.
Exploiting this tunability, one aims to control chemical reactions in a desired manner.
Thus, understanding how time-dependent reaction rates influence non-equilibrium responses and transport is a fundamental problem in stochastic chemical kinetics.

\begin{figure}[!htbp]
  \begin{minipage}{0.48\linewidth}
    \includegraphics[width=\linewidth]{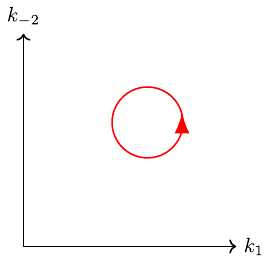}
    \centering
    \label{fig:continuous_scheme}
  \end{minipage}
  \hfill
  \begin{minipage}{0.48\linewidth}
    \includegraphics[width=\linewidth]{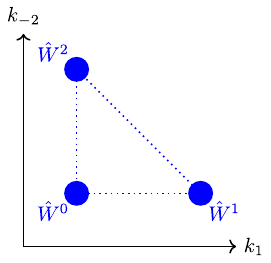}
    \centering
    \label{fig:discreete_scheme}
  \end{minipage}
  \caption{
    Driving protocols schematically shown for (Left) continuous modulation,
    and for (Right) stepwise constant switching of the reaction rates studied in this work.
  }
  \label{fig:driving_protocols}
\end{figure}
\begin{table*}[t!]
  \centering
  \caption{
    Comparison between previous studies and our contribution.
  }
  \label{table:comparison}
  \begin{ruledtabular}
    \begin{tabular}{l c c}
      &
        Previous Studies
      & Our Work \\
      &
      \cite{Astumian1992-qd,Astumian2001-je,Sinitsyn2007-qo,Ohkubo2008-za,Ohkubo2008-xl,Sinitsyn2009-jy,Takahashi2020-nz}
      & \\
      \colrule
        Parameter Change
      &
        Continuous
      &
        Both Continuous and Discrete Switching \\
        Driving Amplitude
      &
        Weak (Perturbative) or General (Two-state)~\cite{Takahashi2020-nz}
      & General (Non-perturbative) \\
      High $\Omega$ & \checkmark & \cref{eq:J_high_freq_general,eq:W_eff_Nstep_1st_order} \\
      Crossover & \checkmark & Numerical \\
      Low $\Omega$ & \checkmark & \cref{eq:J_geom_low,eq:J_dyn_low} \\
    \end{tabular}
  \end{ruledtabular}
\end{table*}

When the time-dependent modulation is periodic in time,
we can gain deeper and more systematic understanding of the nonstationary system with the help of the
Floquet formalism, i.e.,
a temporal version of the Bloch theorem using a Fourier transform in the time variable.
The idea of controlling systems and realizing desired properties by employing time-dependent modulations is dubbed \textit{Floquet engineering}.
One of the paradigmatic classical examples is the Kapitza pendulum, where rapid modulation generates an effective potential that stabilizes an otherwise unstable inverted configuration~\cite{Kapitza1965-ve,Citro2015-qv}.
In quantum systems, a wide range of phenomena have been proposed including controllable band-structures and driven topological phases
\cite{Goldman2014-ek,Eckardt2015-qn,Bukov2015-ux,Mikami2016-yv,Oka2019-oj,Morimoto2023-nz,Oka2009-ky,Lindner2011-gc}.
More recently, Floquet-based approaches have also been systematically developed for classical stochastic processes and dissipative systems~\cite{Higashikawa2018-px,Sato2025-gh}.

The classical stochastic process that we focus on in this work is a chemical reaction model catalyzed by a single enzyme~\cite{Michaelis1913-fi,Robertson1991-pj,Astumian1992-qd,Astumian2001-je,Astumian2005-oi}.
This system is known to realize \textit{stochastic pumping} or \textit{geometric current},
where periodic modulation of reaction rates generates net production even in the absence of bias in the averaged rates~\cite{Sinitsyn2007-qo,Ohkubo2008-za,Ohkubo2008-xl,Sinitsyn2009-jy}.
Previous studies mainly considered perturbatively weak and continuous modulation, as in the left panel in \cref{fig:driving_protocols} and \cref{table:comparison}.
In the case of fast high-frequency driving, the net production has been evaluated based on a formulation in terms of the one-period time-evolution operator.
For the slow and low-frequency driving, it was attributed to geometric effects~\cite{Sinitsyn2007-qo,Sinitsyn2009-jy} similar to the Berry phase effects in quantum systems~\cite{Xiao2010-rw}.

We aim to establish a non-perturbative framework for discrete Floquet engineering and apply this framework to classical stochastic processes.
This direction is motivated by the practical realities of stochastic kinetics;
in experimental settings, control parameters frequently undergo abrupt, stepwise changes---such as in binding events, on-off optical switching, and conformational changes of proteins---rather than smooth oscillations.
Under these conditions, where neither smooth-driving nor weak-amplitude approximations hold, a more robust approach is required.
Only within such a framework can we assess the true stability of geometric currents and elucidate the underlying transport mechanisms across the entire frequency spectrum.

However, previous approaches do not by themselves provide a framework for non-perturbative amplitude stepwise switching, where the protocol is specified by an ordered sequence of Markov generators rather than by a smooth weak modulation.
In this paper, we formulate a counting-field-dependent Floquet theory including such discrete driving and express the long-time current through the effective Floquet generator over one period.
This framework yields formal expression for all frequencies, and explicit asymptotic formulas in the high- and low-frequency regimes, based respectively on the van Vleck effective Floquet generator and the geometric phase~\cite{Aharonov1987-uq,Choutri2002-bf,Cohen2019-dg}.
For a three-step enzymatic reaction protocol, we show numerically and analytically that the current is controlled by the cyclic ordering of the generators and the driving frequency, with asymptotic formulas consistent with the numerical results presented in \cref{sec:application_for_discrete_driving_protocols}.

The remainder of this paper is organized as follows.
\cref{sec:model} introduces the enzymatic reaction model, the counting-field formulation, and the target quantities.
\cref{sec:floquet_theory_counting_statistics} develops the Floquet formulation for counting statistics, derives formal expressions for the long-time current, and then gives its general high- and low-frequency expressions.
\cref{sec:application_for_discrete_driving_protocols} applies this framework to discrete driving protocols with biological motivation, presents numerical results for a three-step protocol, derives the corresponding asymptotic currents (\cref{eq:J_high_freq,eq:J_geom_low_concrete}), and compares them with the numerical results.
\cref{sec:conclusion} concludes the paper with a summary and an outlook.

%% file: sections/2-model.tex

\section{Model} \label{sec:model}
  We consider a single enzyme molecule that catalyzes the conversion of a substrate into a product via the following chemical equation (Michaelis--Menten reaction):
  \begin{align}
      \schemestart
        \ce{E} \+ \ce{S}
        \arrow{<=>[$k_1$][$k_{-1}$]}
        \ce{E-S}
        \arrow{<=>[$k_2$][$k_{-2}$]}
        \ce{E} \+ \ce{P}
      \schemestop
      \label{eq:enzymatic_model}
  \end{align}
  where \ce{E} is an enzyme, \ce{S} is a substrate, \ce{E-S} is a bound complex, and \ce{P} is a product~\cite{Michaelis1913-fi,Qian2002-rv,Sinitsyn2007-qo,Ohkubo2008-za,Ohkubo2008-xl,Sinitsyn2009-jy}.
  This reaction comprises four elementary steps: substrate binding/unbinding $(\ce{E + S} \chemrev \ce{E-S})$ and product formation/rebinding $(\ce{E-S} \chemrev \ce{E + P})$.
  The reaction rates are denoted by $k_1$, $k_{-1}$, $k_2$, and $k_{-2}$, respectively.
  We allow the reaction rates to be time-dependent \((k_{\pm1}(t), k_{\pm2}(t))\)~\cite{Astumian2001-je,Astumian2003-sm,Sinitsyn2007-qo}.
  The number of enzyme molecules is fixed at one ($\#\ce{E} = 1$), while \ce{S} and \ce{P} are treated as reservoirs such that any changes in the amount of \ce{S} and \ce{P} have no effect on the reaction rates.

  \begin{table*}[t!]
    \centering
    \caption{
      Notations in this work.
    }
    \label{table:definition_adopted_in_this_work}
    \begin{ruledtabular}
      \begin{tabular}{l c c c}
        Quantity & Eq. & Definition & Terminology\\
        \colrule
        $n$ && Change in the number of \ce{P} from initial time & counting variable \\
        $\chi$ && Fourier-conjugate variable to the counting variable $n$ & counting field \\
        $\mathcal{Z}(\chi; t)$
        &
        (\ref{eq:Z})
        &
        $\displaystyle
          \braket<S| \TimeEvolutionOp(\chi; t, t_0)\,|\psi(\chi; t_0)>
          \left(
            =
              \sum_{n=-\infty}^{\infty}\sum_{c=1,2}
              p_{n,c}(t) \napier^{\imaginaryi n \chi}
          \right)
        $
        &
        generating function
        \\
        $\mathcal{S}(\chi;t)$
        &
        (\ref{eq:CGF})
        &
        $\displaystyle
          \log \mathcal{Z}(\chi; t)
        $
        &
        cumulant generating function \\
        $N_{\ce{P}}(t)$
        &
        (\ref{eq:N_P})
        &
        $\displaystyle
          \sum_{n=-\infty}^{\infty} n \left( p_{n,1}(t) + p_{n,2}(t) \right)
          \left(= - \imaginaryi \partial_{\chi} \mathcal{S}(\chi; t)\big|_{\chi=0}\right)
        $
        & net production
        \\
        $J(\Omega;t)$
        &
        (\ref{eq:J_t})
        &
        $\displaystyle
        \frac{N_{\ce{P}}(t) - N_{\ce{P}}(t_0)}{t - t_0}
          \left(
          =
          - \imaginaryi\partial_\chi
          \frac{\mathcal{S}(\chi;t) - \mathcal{S}(\chi; t_0)}{t-t_0}\Bigg|_{\chi=0}
          \right)
        $
        &
        net current \\
        $J_{\infty}(\Omega)$
        &
        (\ref{eq:J_infty})
        &
        $\displaystyle\lim_{t\to\infty}J(\Omega;t)$
        &
        long-time current
      \end{tabular}
    \end{ruledtabular}
  \end{table*} %
  Let $p_{n,1}(t)$ and $p_{n,2}(t)$ denote the probabilities of finding the enzyme in the free state \ce{E} and in the bound state \ce{E-S}, respectively, at time $t$,
  where $n$ represents the counting variable that tracks the net change in the number of product molecules \ce{P} from the initial time $t_0$.
  They obey the following master equation:
  \begin{align}
    \left\{
    \begin{aligned}
        \frac{\mathrm{d}}{\mathrm{d} t} p_{n,1}
        &= -\big(k_{1}+k_{-2}\big)\, p_{n,1}
          + k_{-1} p_{n,2}
          + k_{2} p_{(n-1),2}, \\
        \frac{\mathrm{d}}{\mathrm{d} t} p_{n,2}
        &= -\big(k_{-1}+k_{2}\big)\, p_{n,2}
          + k_{1} p_{n,1}
          + k_{-2} p_{(n+1),1}.
    \end{aligned}
    \right.
    \label{eq:master_eq}
  \end{align}
  Here, $n \in \mathbb{Z}$.
  Throughout this work, the enzymatic reaction is modeled as a continuous-time Markov process governed by this master equation.

  Next, we introduce the counting field $\chi$.
  This is a standard auxiliary variable widely employed to evaluate the statistics of various observables in stochastic systems, ranging from mesoscopic transport to chemical kinetics~\cite{Esposito2009-jw}.
  In the context of the present enzymatic reaction, this approach has been adopted in several previous studies~\cite{Zheng2013-fw,Ohkubo2013-wb,Honeychurch2020-zc}.
  The idea is to interpret the master equation \cref{eq:master_eq} as a real space Schr\"{o}dinger equation defined on a lattice with position labeled by counting variable $n$.
  The master equation has lattice-translation symmetry, and thus we can move to the momentum space, with $\chi$ as the conjugate momentum of $n$.
  Accordingly, we cast \cref{eq:master_eq} into the Schr\"odinger equation:
  \begin{align}
    \frac{\mathrm{d}}{\mathrm{d} t}\,\ket|\psi(\chi; t)> =\Markov(\chi; t)\,\ket|\psi(\chi; t)>,
    \label{eq:schrodinger_like_eq}
  \end{align}
  where $\ket|\psi(\chi; t)>$ is a vector with components $P_c(\chi; t) \, (c =1, 2)$, the generating function of the probabilities $p_{n,c}(t)$,
  and $\Markov(\chi; t)$ is a generator that depends on the counting field $\chi$ and time $t$, defined as
  \begin{align}
    \ket|\psi(\chi; t)> &\equiv
      \begin{pmatrix}
      P_1(\chi; t) \\
      P_2(\chi; t)
      \end{pmatrix},
      \,
    P_c(\chi; t) \equiv
      \sum_{n=-\infty}^{\infty} p_{n,c}(t)\,\mathrm{e}^{\mathrm{i} n\chi},
      \label{eq:psi_definition}\\
    \Markov(\chi; t) &\equiv
      \begin{pmatrix}
      -\,k_{1}-k_{-2} & k_{-1}+k_{2}\mathrm{e}^{\mathrm{i}\chi} \\
      k_{1}+k_{-2}\mathrm{e}^{-\mathrm{i}\chi} & -\,k_{-1}-k_{2}
      \end{pmatrix}.
    \label{eq:W_chi_definition}
  \end{align}
  This implies that, when calculating statistics of $n$, the original reaction scheme (\ref{eq:enzymatic_model}) is formally modified as
  \begin{align}
      \schemestart
        \ce{E} \+ \ce{S}
        \arrow{<=>[$k_1$][$k_{-1}$]}
        \ce{E-S}
        \arrow{<=>[$k_2\napier^{\imaginaryi\chi}$][$k_{-2}\napier^{-\imaginaryi\chi}$]}
        \ce{E} \+ \ce{P}.
      \schemestop
      \label{eq:enzymatic_model_chi}
  \end{align}

  The $\chi$-dependent propagator is defined as
  \begin{align}
  \TimeEvolutionOp(\chi; t_2,t_1)
  &=
    \mathcal{T}\exp\Bigl(\int_{t_1}^{t_2}\! \Markov(\chi; s)\,\mathrm{d} s\Bigr),
  \label{eq:U_chi}
  \end{align}
  where $\mathcal{T}$ is a time-ordered product.
  The time evolution of the state vector is given by
  \begin{align}
    \ket|\psi(\chi; t)> =
      \TimeEvolutionOp(\chi; t, t_0)\,\ket|\psi(\chi; t_0)>.
    \label{eq:psi_t}
  \end{align}
  Let us consider a relation derived from the probability conservation rules for later analytical calculations.
  For all time $t$, the total probability is conserved: $P_1(0;t) + P_2(0;t) = \braket<S|\psi(0;t)>=1$, with $\bra<S| \equiv \begin{pmatrix}1&1\end{pmatrix}$ being the \textit{projection state}, and this provides a restriction on the generator $\Markov(0;t)$:
  \begin{align}
    \frac{\mathrm{d}}{\mathrm{d}t}\braket<S|\psi(0;t)>
    &= \bra<S|\,\Markov(0;t)\,\ket|\psi(0;t)>=0, \nonumber \\
    \therefore \bra<S|\,\Markov(0;t)&=0.
    \label{eq:probability_conservation}
  \end{align}
  This means that $\bra<S|$ is a left zero-eigenvector of $\Markov(0;t)$ for all $t$.
  Note that in general, $\bra<S|$ is not a left eigenvector of $\Markov(\chi;t)$ for $\chi \neq 0$.
  When we use the consequence of probability conservation, we always have to set $\chi = 0$.

  In order to evaluate the statistics of the counting variable $n$, we use the generating function $\mathcal{Z}(\chi;t)$ and the cumulant generating function $\mathcal{S}(\chi;t)$ \cite{Sinitsyn2007-qo}:
  \begin{align}
      \mathcal{Z}(\chi; t) &\equiv \braket<S| \TimeEvolutionOp(\chi; t, t_0)\,|\psi(\chi; t_0)>,
      \label{eq:Z} \\
      \mathcal{S}(\chi;t) &\equiv \log \mathcal{Z}(\chi; t).
      \label{eq:CGF}
  \end{align}
  Derivatives of the cumulant generating function at ${\chi=0}$ yield the time-dependent statistics of the counting variable $n$, i.e.,
  the $l$-th derivative in terms of $\mathrm{i} \chi$ gives the $l$-th cumulant of $n$.
  For example, the first derivative gives the expectation value or the net production as follows:
  \begin{align}
    N_{\ce{P}}(t)
    &=
        \sum_{n=-\infty}^{\infty} n \left( p_{n,1}(t) + p_{n,2}(t) \right)
    =
    - \imaginaryi \partial_{\chi} \mathcal{S}(\chi;t)|_{\chi=0}.
  \label{eq:N_P}
  \end{align}
  We denote the corresponding time-averaged current by $J(\Omega; t)$, and refer to it as the net current.
  It is given by
  \begin{align}
      J(\Omega; t) &\equiv
        \frac{N_{\ce{P}}(t) - N_{\ce{P}}(t_0)}{t-t_0},
      \label{eq:J_t}\\
      J_\infty(\Omega) &\equiv
        \lim_{t\to\infty} J(\Omega; t).
      \label{eq:J_infty}
  \end{align}
  We refer to \cref{eq:J_infty} as the \textit{long-time current}.
  For later convenience, we summarize the definitions of the quantities introduced in this section in \cref{table:definition_adopted_in_this_work}.

%% file: sections/3-floquet-theory.tex

\section{Floquet Theory for Counting Statistics in Periodically Driven Markov Processes}
\label{sec:floquet_theory_counting_statistics}

We now formulate the Floquet theory for a periodically driven continuous-time Markov process in the presence of a counting field.
Although the counting-field formalism introduced in \cref{sec:model} can be generalized to multiple counting fields, allowing cumulants and correlations of arbitrary order to be calculated~\cite{del-Razo2026-vs}, we focus here on a single counting field and the first cumulant.
This restriction is essential in extracting the physical consequences of periodic driving most clearly.
Our goal in this section is to derive general expressions for the long-time current in two asymptotic regimes: the high-frequency regime, where the dynamics is governed by an effective Floquet generator, and the low-frequency regime, where the system relaxes within each slowly varying time interval.

\subsection{Floquet Representation with a Counting Field}

Let us consider a continuous-time Markov process whose counting-field-dependent generator is periodic in time,
\begin{align}
  \Markov(\chi;t+T)=\Markov(\chi;t),
  \qquad
  \Omega=\frac{2\pi}{T}.
\end{align}
Here, $T$ is a period of driving.
The time-evolution operator is
\begin{align}
  \TimeEvolutionOp(\chi;t,t_0)
  =
  \mathcal{T}
  \exp\!\left(
    \int_{t_0}^{t}
    \Markov(\chi;s)\,\mathrm{d}s
  \right).
  \label{eq:U_chi_t_t0}
\end{align}
Floquet theory~\cite{Shirley1965-nc,Shavitt1980-tx,Goldman2014-ek,Eckardt2015-qn,Mikami2016-yv,Eckardt2017-sm,Morimoto2023-nz} allows us to decompose this operator as
\begin{align}
  \TimeEvolutionOp(\chi;t,t_0)
  =
  \hat{V}(\chi;t)\,
  \napier^{\Markov^{\mathrm{eff}}(\chi)\, (t - t_0)}
  \hat{V}^{-1}(\chi;t_0),
  \label{eq:U_Floquet_tripartitioned_chi}
\end{align}
where $\hat{V}(\chi;t+T)=\hat{V}(\chi;t)$ is the micromotion operator and $\Markov^{\mathrm{eff}}(\chi)$ is the effective Floquet generator.
The right and left Floquet states form a biorthogonal basis.
At $\chi=0$, probability conservation gives
\begin{align}
  \bra<S|\,\Markov(0;t)=0,
  \qquad
  \bra<S|\,\TimeEvolutionOp(0;t,t_0)=\bra<S|.
  \label{eq:probability_conservation_floquet}
\end{align}

The generating function $\mathcal{Z}(\chi;t)$, the cumulant generating function $\mathcal{S}(\chi;t)$, and the net production $N_{\ce{P}}(t)$ have already been introduced in \cref{eq:Z,eq:CGF,eq:N_P}.
Let us denote $\Delta t = t - t_0$.
The counting-field derivative of $\napier^{\Markov^{\mathrm{eff}}(\chi)\Delta t}$ cannot be replaced by $-\imaginaryi \,\Delta t \,\partial_{\chi}\Markov^{\mathrm{eff}}(\chi) \cdot \napier^{\Markov^{\mathrm{eff}}(\chi)\Delta t}$, because $\Markov^{\mathrm{eff}}(\chi)$ and $\partial_\chi\Markov^{\mathrm{eff}}(\chi)$ are generally non-commutative.
Instead, we introduce the following operator
\begin{align}
    \hat{\mathcal{J}}^{\mathrm{eff}}_{\Delta t}(\chi)
    \equiv
    - \frac{\imaginaryi}{\Delta t} \partial_{\chi}
    \left(\napier^{\Markov^{\mathrm{eff}}(\chi) \Delta t}\right)\,
    \napier^{-\Markov^{\mathrm{eff}}(\chi) \Delta t}.
    \label{eq:time_averaged_current_op}
\end{align}
Using \cref{eq:Duhamel_formula}, this operator can be rewritten as
\begin{align}
    \hat{\mathcal{J}}^{\mathrm{eff}}_{\Delta t}(\chi)
    &=
    \frac{1}{\Delta t}
    \int_0^{\Delta t} \mathrm{d}s\,
    \napier^{\Markov^{\mathrm{eff}}(\chi)s}
    \hat{J}^{\mathrm{eff}}(\chi)
    \napier^{-\Markov^{\mathrm{eff}}(\chi)s},
    \\
    \hat{J}^{\mathrm{eff}}(\chi)
    &\equiv
    -\imaginaryi\partial_{\chi}\Markov^{\mathrm{eff}}(\chi).
    \label{eq:finite_time_effective_current_operator_average}
\end{align}
Thus, $\hat{\mathcal{J}}^{\mathrm{eff}}_{\Delta t}(\chi)$ is the time average of the effective current operator $\hat{J}^{\mathrm{eff}}(\chi)$ in the interaction representation associated with $\Markov^{\mathrm{eff}}(\chi)$.
Using this operator, we can formally rewrite the net production as
\begin{align}
    N_{\ce{P}}(t) =
    \Delta t \, \braket*<S|\hat{\mathcal{J}}^{\mathrm{eff}}_{\Delta t}(\chi)
    |\Psi(\chi;t)>\Big|_{\chi=0}
    +\, \mathrm{residual\ terms}.
\end{align}
Here, we used the three identities:\cref{eq:probability_conservation_floquet}, $\mathcal{Z}(\chi;t)|_{\chi=0}=1$, and ${\hat{V}^{-1}(\chi;t)\hat{V}(\chi;t)} = \hat{I}$.
The state $\ket|\Psi(\chi;t)>$ is defined by
\begin{align}
    \ket|\Psi(\chi;t)>
    &\equiv
    \napier^{\Markov^{\mathrm{eff}}(\chi )\Delta t} \hat{V}^{-1}(\chi;t_0)\ket|\psi(\chi;t_0)> \nonumber \\
    &=
    \hat{V}^{-1}(\chi;t) \ket|\psi(\chi;t)>,
\end{align}
which we refer to as the \textit{kicked state}.
Thus, we obtain the formal expression below:
\begin{align}
    N_{\ce{P}}(t)&= \Delta t \cdot J(\Omega;t) + \mathrm{residual\ terms},
    \label{eq:N_P_low_freq_formal}
    \\
    J(\Omega;t) &= \braket<S|\hat{\mathcal{J}}^{\mathrm{eff}}_{\Delta t}(0)|\Psi(0;t)>.
    \label{eq:J_formal}
\end{align}
This expression shows that the net current at time $t$ is determined by the finite-time effective current operator $\hat{\mathcal{J}}^{\mathrm{eff}}_{\Delta t}(\chi)$ and the kicked state $\ket|\Psi(\chi;t)>$.
Nevertheless, their explicit forms remain unknown at this stage.
To obtain computable expressions, we must evaluate $\Markov^{\mathrm{eff}}(\chi)$, $\hat{\mathcal{J}}^{\mathrm{eff}}_{\Delta t}(\chi)$, together with $\ket|\Psi(0;t)>$.
In the following sections, we do this in two frequency regimes.
In the low-frequency regime, a geometric pumping originating from geometric phase contributes to the current.
In the high-frequency regime, the current is characterized by the effective Floquet generator.
The behavior at an arbitrary time can be obtained directly from \cref{eq:J_formal} by tracking the continuous time evolution.
Here, however, we further simplify the calculation by setting the final time to $t = t_0 + M T$, where $M$ is a positive integer.
This allows us to focus on the stroboscopic behavior of the current, while also leading to a simple analytic expression.

\subsection{High-Frequency Regime}

In the high-frequency regime, the generator changes so rapidly that the state cannot follow the instantaneous steady state.
Instead, the state and the long-time current are governed by the effective Floquet generator that describes the dynamics averaged over one period.
We use the van Vleck high-frequency expansion to capture the effective dynamics and the micromotion separately.
This allows us to extract the long-time current in a systematic expansion in powers of $\Omega^{-1}$.
With the Fourier expansion
\begin{align}
  \Markov(\chi;t)
  =
  \sum_{m=-\infty}^{\infty}
  \Markov_m(\chi)\,
  \napier^{-\imaginaryi m\Omega t},
\end{align}
the effective Floquet generator and the kick operator up to the first order in $\Omega^{-1}$ are~\cite{Kitagawa2011-ia}
\begin{align}
  \Markov^{\mathrm{eff,vV}}(\chi)
  &=
  \Markov_0(\chi)
  +
  \imaginaryi
  \sum_{m\neq0}
  \frac{
    [\Markov_{-m}(\chi),\Markov_m(\chi)]
  }{2m\Omega}
  +
  \mathcal{O}(\Omega^{-2}),
  \label{eq:Markov_eff_vV_general}
  \\
  \KickMarkov^{\mathrm{vV}}(\chi;t)
  &=
  \imaginaryi
  \sum_{m\neq0}
  \frac{\Markov_m(\chi)}{m\Omega}
  \napier^{-\imaginaryi m\Omega t}
  +
  \mathcal{O}(\Omega^{-2}),
  \label{eq:KickMarkov_vV_general}
  \\
  \hat{V}(\chi;t)
  &=
  \napier^{\KickMarkov^{\mathrm{vV}}(\chi;t)}.
\end{align}
See \cref{app:high_freq_expansion} for the derivation.
At $\chi=0$, probability conservation implies $\bra<S|\Markov_m(0)=0$ for all $m$, and therefore
\begin{align}
  \bra<S|\,\Markov^{\mathrm{eff,vV}}(0)=0,
  \qquad
  \bra<S|\,\KickMarkov^{\mathrm{vV}}(0;t)=0.
  \label{eq:probability_conservation_vV}
\end{align}

We can see that, in the long-time limit with stroboscopic times, \cref{eq:J_formal} reduces to
\begin{align}
    J_{\infty}(\Omega) &=
        \braket<S|\hat{\mathcal{J}}^{\mathrm{eff}}_{\Delta t}(0)|\Psi(0;t)> = \braket<S|\hat{J}^{\mathrm{eff}}(0)|\psi_0>,
    \label{eq:J_eff}
\end{align}
where
\begin{align}
    \hat{J}^{\mathrm{eff}}(\chi) &=
        - \imaginaryi \partial_\chi \Markov^{\mathrm{eff}}(\chi).
\end{align}
The rigorous manipulations are rather involved; we therefore defer the details to \cref{app:rigorous_derivation_of_J_high}.
By inserting \cref{eq:Markov_eff_vV_general} into \cref{eq:J_eff} and appropriately decomposing the rotating-frame contribution in the long-time limit, we obtain
\begin{align}
  J_{\infty}(\Omega)
  &=
  \braket<S|
  \left(
    -\imaginaryi
    \partial_{\chi}
    \Markov^{\mathrm{eff,vV}}(\chi)
    \Big|_{\chi=0}
  \right)
  |\psi_0>
  \nonumber \\
  &=
  \braket<S|
  \left(
    \hat{J}_0
    +
    \imaginaryi
    \sum_{m\neq0}
    \frac{[\Markov_{-m},\hat{J}_m]}{m\Omega}
    +
    \mathcal{O}(\Omega^{-2})
  \right)
  |\psi_0>,
  \label{eq:J_high_freq_general}
\end{align}
where $\ket|\psi_0>$ is the normalized right zero-eigenvector of $\Markov^{\mathrm{eff,vV}}(0)$.

The central role of the present framework is that it provides a systematic high-frequency expansion of the effective Floquet generator beyond the leading order in $\Omega^{-1}$.
This structure has two important consequences.
First, higher-order terms can be included explicitly to improve the accuracy of analytic calculations.
As a result, the present formulation is not restricted to the asymptotic high-frequency regime, but can systematically improve analytic approximations toward intermediate-frequencies as long as the expansion remains controlled.

Second, the framework is expected to be useful even when the leading correction does not appear at order $\Omega^{-1}$.
Indeed, in some periodically driven systems, including the Kapitza pendulum and related Floquet systems~\cite{Bukov2015-ux,Citro2015-qv}, the first nonzero correction arises at order $\Omega^{-2}$.
Such cases require a formulation that can
retain higher-order Floquet corrections systematically.

\subsection{Low-Frequency Regime}

In the low-frequency regime, the generator typically changes slowly compared with the relaxation time of the Markov process.
In this section, we briefly review the approach introduced in Ref.~\cite{Sinitsyn2007-qo}, which was applied for a continuously driven system.
In the next section \cref{sec:application_for_discrete_driving_protocols}, we extend it to a discretely driven system.
We divide the time interval $[t_0, t]$ into $L(=MN)$ intervals, $t_j=t_0+j\delta t$ with $t_L=t$.
$N$ represents the number of intervals within one period.
Following the argument in Ref.~\cite{Sinitsyn2007-qo}, we characterize this $\delta t$ as an intermediate time scale $\delta t$ satisfying
\begin{align}
  T \gg \delta t \gg \tau_{\mathrm{rel}},
  \label{eq:intermediate_time_scale_low_freq}
\end{align}
where $\tau_{\mathrm{rel}}$ is the typical relaxation time of the instantaneous generator.
For a stepwise protocol, this condition should be understood within each dwell interval: each interval must be long compared with the relaxation time of the corresponding generator.

For each interval, the generator can be regarded as approximately constant, and the generating function is written as
\begin{align}
  \mathcal{Z}(\chi;t)
  \simeq
  \braket<S|
    \napier^{\Markov(\chi;t_L)\delta t} \cdots
    \napier^{\Markov(\chi;t_1)\delta t}
  |\psi(\chi;t_0)>.
  \label{eq:Z_low_freq_discretized}
\end{align}
We now insert the instantaneous biorthogonal resolution of identity,
\begin{align}
  \hat{I}
  =
  \sum_{s=0}^{D-1}
  \ketbra*|u_s(\chi;t_j)><\tilde{u}_s(\chi;t_j)|,
\end{align}
where $D$ is the dimension of the generator $\Markov(\chi;t_j)$, and $\ket|u_s(\chi;t_j)>, \bra<\tilde{u}_s(\chi;t_j)|$ are the right and left eigenmodes of $\Markov(\chi;t_j)$.
Let $\lambda_0(\chi;t_j)$ denote the eigenvalue with the largest real part, which we refer to as the \textit{dominant eigenvalue} and the corresponding right and left eigenvectors as the \textit{dominant eigenmodes}.
At $\chi=0$, the Perron--Frobenius theorem guarantees that the zero mode of an irreducible Markov generator is simple and that all other eigenvalues have negative real parts (see \cref{app:Perron_Frobenius_Theorem}).
Therefore, for $\chi$ around zero, the dominant eigenvalue and the corresponding right and left eigenvectors can be followed continuously.
Since $\delta t$ is longer than the relaxation time, the contribution from this dominant mode remains after each interval, while the other modes are exponentially suppressed.
For smoothly driven Markov processes, systematic treatments based on slow-driving expansion and the improved adiabatic approximation have also been developed~\cite{Mandal2016-sz,Matsuo2022-nl}.

After the repeated projection onto this mode, the generating function becomes
\begin{align}
  \mathcal{Z}(\chi;t)
  &\simeq
  \braket<S|u_0(\chi;t_L)> \nonumber \\
  &\quad\cdot
  \prod_{j=1}^{L}
  \left[
    \braket*<\tilde{u}_0(\chi;t_j)|u_0(\chi;t_{j-1})>
    \napier^{\lambda_0(\chi;t_j)\delta t}
  \right]
  \nonumber \\
  &\quad
  \cdot
  \braket<\tilde{u}_0(\chi;t_0)|\psi(\chi;t_0)>.
  \label{eq:Z_low_freq_dominant_mode}
\end{align}
We denote
\begin{align}
  \mathcal{Z}_{\mathrm{geom}}(\chi)
  &=
  \prod_{j=1}^{N}
  \braket*<\tilde{u}_0(\chi;t_j)|u_0(\chi;t_{j-1})>,
  \label{eq:Z_geom_low_freq}
  \\
  \mathcal{Z}_{\mathrm{dyn}}(\chi)
  &=
  \prod_{j=1}^{N}
  \napier^{\lambda_0(\chi;t_j)\delta t},
  \label{eq:Z_dyn_low_freq}
  \\
  \mathcal{Z}_{\mathrm{i+f}}(\chi;t,t_0)
  &=
    \braket<S|u_0(\chi;t)>
    \braket<\tilde{u}_0(\chi;t_0)|\psi(\chi;t_0)>,
  \label{eq:Z_if_low_freq}
\end{align}
to decompose $\mathcal{Z}(\chi;t)$ as
\begin{align}
  \mathcal{Z}  
    =
    \mathcal{Z}_{\mathrm{geom}}^M
    \mathcal{Z}_{\mathrm{dyn}}^M
    \mathcal{Z}_{\mathrm{i+f}}.
    \label{eq:Z_decomposed_low_freq}
\end{align}
The three factors correspond to the geometric, dynamical, and initial--final contributions, respectively.
Using \cref{eq:Z_decomposed_low_freq}, the net production \cref{eq:N_P} is written as
\begin{widetext}
\begin{align}
    N_{\ce{P}}(t_0+MT)
    &=
        MT \cdot \frac{-\imaginaryi\partial_{\chi}}{T} \left(\mathcal{S}_{\mathrm{geom}} + \mathcal{S}_{\mathrm{dyn}}
        \right)\Big|_{\chi=0}  - \imaginaryi\partial_{\chi} \mathcal{S}_{\mathrm{i+f}}\Big|_{\chi=0},
    \label{eq:N_P_t0MT} \\
    \mathcal{S}_{\mathrm{geom}} &= \log\mathcal{Z}_{\mathrm{geom}}, \quad \mathcal{S}_{\mathrm{dyn}} = \log\mathcal{Z}_{\mathrm{dyn}}, \quad \mathcal{S}_{\mathrm{i+f}} = \log\mathcal{Z}_{i+f},
\end{align}
\end{widetext}
where $- T^{-1} \imaginaryi \partial_{\chi} (\mathcal{S}_{\mathrm{geom}} + \mathcal{S}_{\mathrm{dyn})}|_{\chi=0}$ corresponds to \cref{eq:J_formal}.
The first term grows linearly in time, whereas the second term gives transient and oscillatory contributions.
This follows from the definitions in \cref{eq:Z_geom_low_freq,eq:Z_dyn_low_freq,eq:Z_if_low_freq}: $-T^{-1}\imaginaryi\partial_{\chi}(\log\mathcal{Z}_{\mathrm{geom}}(\chi)\mathcal{Z}_{\mathrm{dyn}}(\chi))$ is independent of $t$, while $-\imaginaryi\partial_{\chi}\log\mathcal{Z}_{\mathrm{i+f}}(\chi;t,t_0)$ is $T$-periodic in $t$.
By comparing \cref{eq:N_P_t0MT} with \cref{eq:N_P_low_freq_formal}, we obtain the explicit form of the long-time current as
\begin{align}
  J_{\infty}(\Omega)
  &=
  J_{\mathrm{geom}}(\Omega)
  +
  J_{\mathrm{dyn}},
  \label{eq:J_low}
  \\
  J_{\mathrm{geom}}(\Omega)
  &=
  \frac{\Omega}{2\pi}
  \partial_{\chi}
  \ImaginaryPartOf
  \log
  \prod_{j=1}^{N}
  \braket*<\tilde{u}_0(\chi;t_j)|u_0(\chi;t_{j-1})>
  \Big|_{\chi=0},
  \label{eq:J_geom_low_freq_general}
  \\
  J_{\mathrm{dyn}}
  &=
  \sum_{j=1}^{N}
  -\imaginaryi\partial_{\chi}
  \lambda_0(\chi;t_j)
  \Big|_{\chi=0}\frac{1}{N}.
  \label{eq:J_dyn_low_freq_general}
\end{align}
Here, we have replaced the expression $-\imaginaryi\partial_{\chi}\log *|_{\chi=0}$ by $\partial_{\chi}\ImaginaryPartOf\log *|_{\chi=0}$ in order to emphasize that the current is real-valued; see \cref{app:pure_imaginary_nature_of_partial_chi_log_Gamma} for details.
This expression shows that the geometric contribution is proportional to the driving frequency $\Omega$, whereas the dynamical contribution has no explicit $\Omega$ dependence at this order.
The geometric part is determined by the sequence of instantaneous dominant eigenvectors rather than by the detailed transient dynamics inside each interval.
The factor $1/N$ in $J_{\mathrm{dyn}}$ is the uniform-dwell-time weight assigned to each step.

\subsection{Remark}
The expressions derived in this section do not depend on the details of the periodic protocol.
In particular, they can be applied not only to protocols that vary smoothly and perturbatively, but also to protocols that switch discretely.
Such discretely switched protocols are also well suited to discussing how discrete external control can influence reaction processes such as the enzymatic model defined in \cref{sec:model}.
In the next section, we therefore apply the present framework to stepwise-constant driving sequences and compare the resulting analytical expressions with direct numerical calculations for the enzymatic reaction model.

%% file: sections/4-application-for-discrete-driving-protocols.tex

\section{Application to Discrete Driving Protocols}
\label{sec:application_for_discrete_driving_protocols}

\begin{figure}[t]
    \centering
    \includegraphics[width=\linewidth]{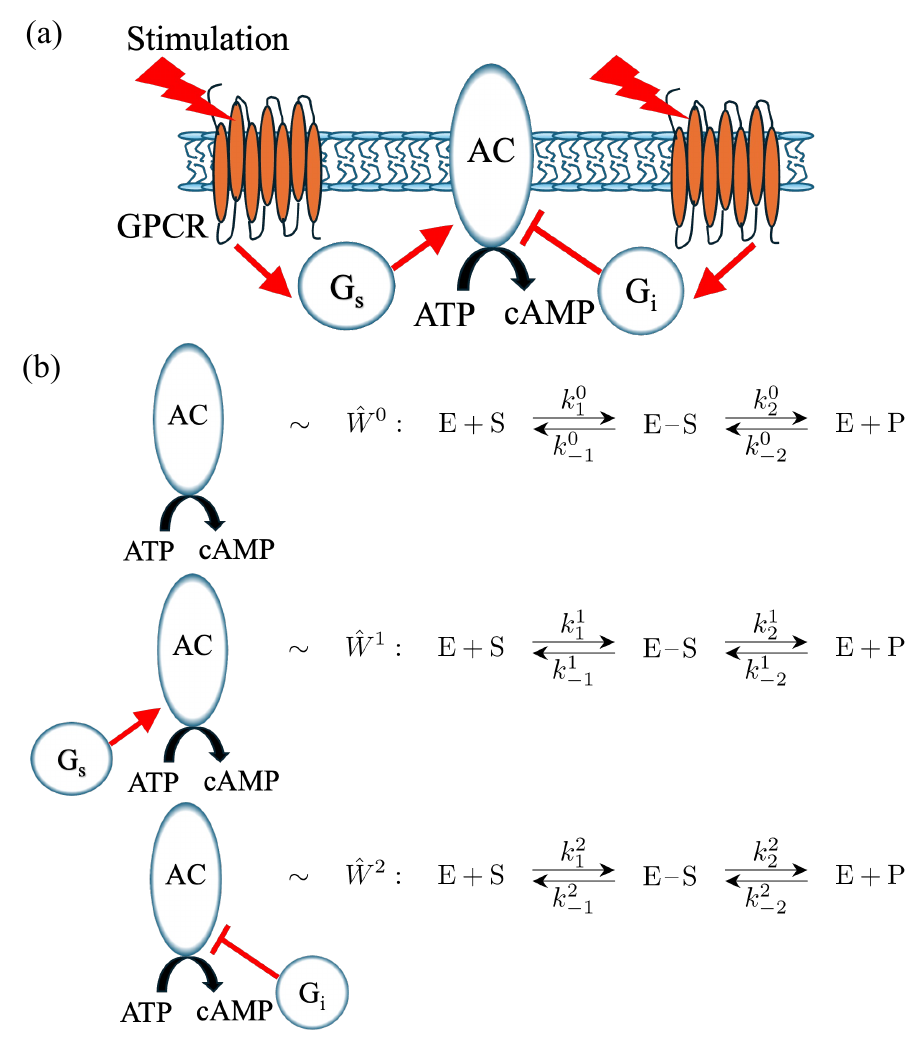}
    \caption{
        Schematic of the cAMP production system motivating discrete Floquet driving.
        (a)
        AC catalyzes the conversion of ATP into cAMP, which correspond to the enzyme E, substrate S, and product P, respectively, in the Michaelis--Menten-type reaction scheme (\ref{eq:enzymatic_model}).
        The regulatory proteins \Gs and \Gi modulate the activation state of AC and thereby alter the catalytic rates in response to external stimulation of GPCRs.
        (b)
        AC-state-dependent chemical reactions, together with the corresponding assumed generators and chemical reaction rates introduced in \cref{eq:rate_constant0,eq:rate_constant1,eq:rate_constant2}.
    }
    \label{fig:cAMP_production_system}
  \end{figure}

Our goal is to investigate the applicability of the Floquet theory developed in \cref{sec:floquet_theory_counting_statistics} to continuous-time Markov processes under discretely switched driving.
We apply the general results of \cref{sec:floquet_theory_counting_statistics} to the enzymatic reaction model introduced in \cref{sec:model} and compare the analytical results with numerical calculations.
Before specifying the driving protocol, we first describe the biological setting that motivates state-dependent rate switching.
In biochemical reactions, the reaction rates often change not gradually but in abrupt jumps, because the proteins involved occupy discrete states---bound versus unbound, or active versus inactive conformations---with a distinct rate in each.
A representative example is the production of cyclic adenosine monophosphate (cAMP) as depicted in \cref{fig:cAMP_production_system}: The enzyme adenylyl cyclase (AC) catalyzes the conversion of ATP into cAMP.
In this system, AC, ATP and cAMP play the roles of E, S, and P in the Michaelis–Menten reaction scheme
(\ref{eq:enzymatic_model}).
The regulatory proteins \Gs and \Gi
switch AC between activated and inhibited states~\cite{bib:Massimi2019-cb}. 
Since AC occupies discrete activation states, its catalytic rates take correspondingly discrete values, so the modulation is naturally stepwise rather than weak and continuous. Optogenetic control of such state-switching proteins, including G-protein-coupled receptors (GPCR)~\cite{Nagata2021-pp,Sakai2022-ws,Boyden2005-xn,Copits2021-mg}, allows these states to be toggled externally and periodically in time.
For more realistic situations,
kinetic information from detailed biochemical studies~\cite{Dessauer1997-fy,Chen-Goodspeed2005-mi} would be helpful for calibrating the reaction rates in optogenetically controlled cAMP production systems.
Motivated by this setting, we adopt the Michaelis--Menten reaction of \cref{sec:model} as a minimal model and impose periodic, stepwise changes on its rate constants, formulating the dynamics as a time-periodic Markov generator.

  Motivated by these settings, we consider the \textit{discrete Floquet driving protocol} below.
  We divide one period into $N$ uniform subintervals,
  \begin{align}
    I_a
    &\equiv
    \{\, t \mid (t \bmod T)\in [a\tau,(a+1)\tau) \,\},
    \nonumber \\
    \tau
    &\equiv
    \frac{T}{N},
    \qquad
    a=0,1,\ldots,N-1.
    \label{eq:discrete_step_intervals}
  \end{align}
  On each interval $I_a$, the rate constants are held fixed.
  In the present enzymatic reaction model, the generator \cref{eq:W_chi_definition} on the interval $I_a$ is denoted by $\Mstep{a}(\chi)$ and is specified by
  \begin{align}
    \bigl(\kstep{a}_{1},\,\kstep{a}_{-1},\,\kstep{a}_{2},\,\kstep{a}_{-2}\bigr).
  \end{align}
  Since $\Markov(\chi;t)=\Mstep{a}(\chi)$ for $t\in I_a$, the one-period propagator is
  \begin{align}
    \TimeEvolutionOp(\chi;T,0)
    =
    \napier^{\Mstep{N-1}(\chi)\tau}
    \cdots
    \napier^{\Mstep{1}(\chi)\tau}
    \napier^{\Mstep{0}(\chi)\tau}.
    \label{eq:discrete_one_period_propagator}
  \end{align}
  We denote this ordered sequence as
  \begin{align}
    \gamma
    =
    \{\Mstep{0},\Mstep{1},\ldots,\Mstep{N-1}\}.
  \end{align}
  This framework originates from the bang-bang protocol~\cite{Lapidus1967-pi,Takahashi1970-zv,Viola1998-ci,Viola1999-zf}, a stepwise control technique developed in optimal control and later applied to quantum systems such as cold atomic systems~\cite{Biercuk2009-ra,Sagi2010-sl}.
  In contrast, our study focuses on classical stochastic processes, where the generator describing the dynamics is generally non-Hermitian.
  The $N$-step formulation has a natural continuous-driving limit: an arbitrary continuous periodic driving protocol can be approximated in the limit $N\to\infty$ by choosing the generators $\Mstep{a}(\chi)$ appropriately.

\begin{figure}[t]
    \begin{minipage}[t]{0.24\textwidth}
      \centering
      \includegraphics[width=\linewidth]{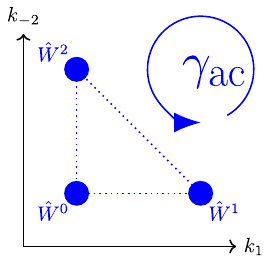}
    \end{minipage}\hfill
    \begin{minipage}[t]{0.24\textwidth}
      \centering
      \includegraphics[width=\linewidth]{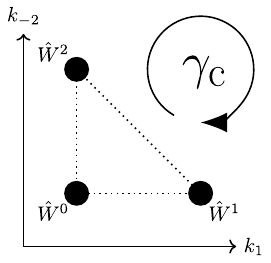}
    \end{minipage}
    \caption{
      Schematic representation of the driving sequences.
      (Left) Anti-clockwise driving sequence $\gammaAntiClockwise = \{\Mstep{0}, \Mstep{1}, \Mstep{2}\}$.
      (Right) Clockwise driving sequence $\gammaClockwise = \{\Mstep{0}, \Mstep{2}, \Mstep{1}\}$.
      Each generator $\Mstep{a}$ is represented by a node, and changes to the next generator every one third of the period $T$.
    }
    \label{fig:driving_sequences}
  \end{figure}

  In the following, we focus on a concrete example and compare numerical and analytical results.
  As an example, we consider the following three-step protocol:
  \begin{align}
    t \in I_0:&\quad
    (k_1^0, k_{-1}^0, k_2^0, k_{-2}^0)
    =
    (1, 1, 1, 1),
    \label{eq:rate_constant0}
    \\
    t \in I_1:&\quad
    (k_1^1, k_{-1}^1, k_2^1, k_{-2}^1)
    =
    (2, 1, 1, 1),
    \label{eq:rate_constant1}
    \\
    t \in I_2:&\quad
    (k_1^2, k_{-1}^2, k_2^2, k_{-2}^2)
    =
    (1, 1, 1, 2).
    \label{eq:rate_constant2}
  \end{align}
  We denote this driving sequence as $\gammaAntiClockwise = \{\Mstep{0}, \Mstep{1}, \Mstep{2}\}$ and its reversed sequence as $\gammaClockwise = \{\Mstep{0}, \Mstep{2}, \Mstep{1}\}$ (\cref{fig:driving_sequences}).
  These two protocols use the same generators with the same duration and differ only in their cyclic ordering.
  The parameters in \cref{eq:rate_constant0,eq:rate_constant1,eq:rate_constant2} are selected such that the dynamical contribution \cref{eq:J_dyn_low_freq_general} vanishes.
  A three-step cycle is the minimal nontrivial choice, since a two-step cycle has no distinct reversed ordering once a cyclic shift of the starting point is ignored.

  To compute the net production $N_{\ce{P}}(t)$, we numerically solve the following time-evolution equations:
  \begin{align}
    \frac{\mathrm{d}}{\mathrm{d} t} \ket*|\psi^{(0)}(t)>
    &=
    \Markov^{(0)}(t) \ket*|\psi^{(0)}(t)>,
    \label{eq:DE_new_1}\\
    \frac{\mathrm{d}}{\mathrm{d} t} \ket*|\psi^{(1)}(t)>
    &=
    \Markov^{(0)}(t) \ket*|\psi^{(1)}(t)>
    +
    \Markov^{(1)}(t) \ket*|\psi^{(0)}(t)>.
    \label{eq:DE_new_2}
  \end{align}
  These equations are obtained by expanding \cref{eq:schrodinger_like_eq} around $\chi=0$ and comparing the coefficients of $\chi^0$ and $\chi^1$ on both sides.
  Here,
  \begin{align}
    \ket*|\psi^{(0)}(t)>
    &\equiv
    \ket|\psi(\chi;t)> \Big|_{\chi=0},
    \quad
    \ket*|\psi^{(1)}(t)>
    \equiv
    \partial_{\chi} \ket|\psi(\chi;t)> \Big|_{\chi=0},
    \\
    \Markov^{(0)}(t)
    &\equiv
    \Markov(\chi;t) \Big|_{\chi=0},
    \quad
    \Markov^{(1)}(t)
    \equiv
    \partial_{\chi}\Markov(\chi;t) \Big|_{\chi=0}.
  \end{align}
  Using the solution of \cref{eq:DE_new_1,eq:DE_new_2}, the net production is computed as
  \begin{align}
    N_{\ce{P}}(t)
    =
    -\imaginaryi
    \braket<S|\psi^{(1)}(t)>.
  \end{align}

\begin{figure*}[t]
    \begin{minipage}[t]{0.65\textwidth}
      \centering
      \includegraphics[width=\linewidth]{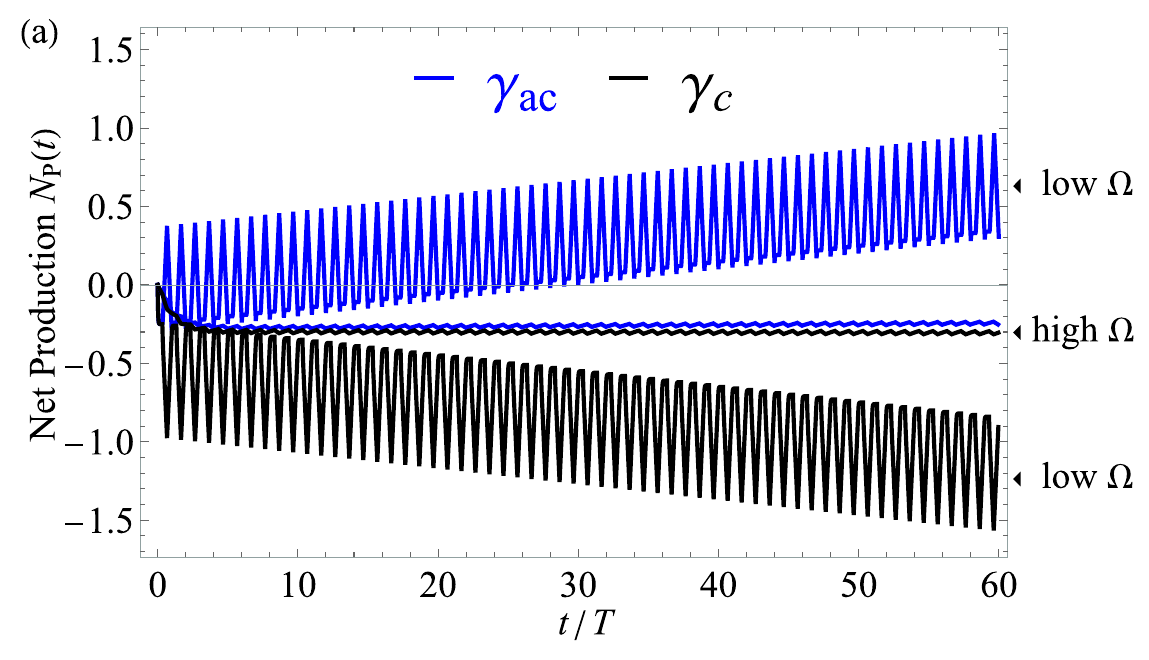}
    \end{minipage}
    \begin{minipage}[t]{\textwidth}
      \centering
      \includegraphics[width=0.48\linewidth]{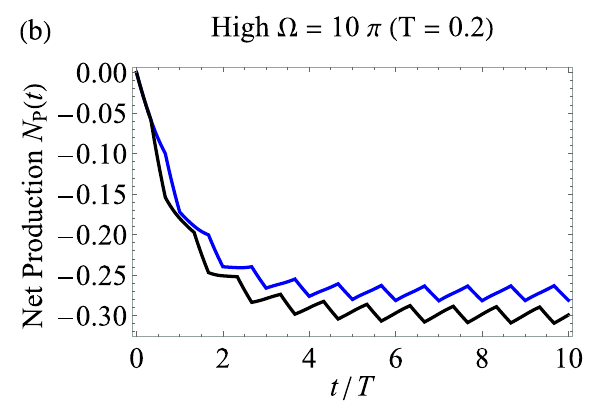}
      \hfill
      \includegraphics[width=0.48\linewidth]{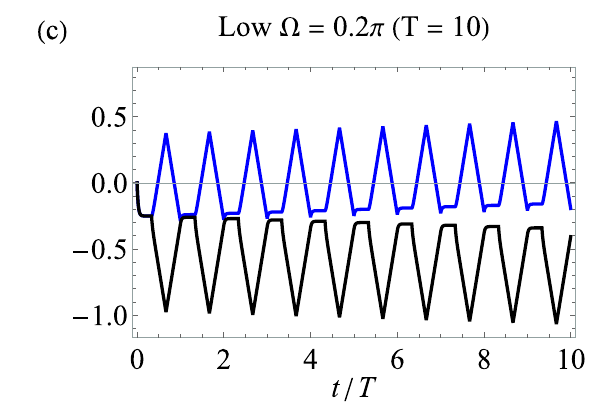}      
    \end{minipage}
    \caption{
      (a) $N_{\ce{P}}(t)$ for different driving sequences and frequencies.
      Blue and black curves correspond to anti-clockwise and clockwise sequences $\gammaAntiClockwise$ and $\gammaClockwise$, respectively.
      (b,c) Enlarged views in the high- and low-frequency regimes.
      We take $\Omega = 10 \pi$ and $\Omega = 0.2 \pi$, respectively.
    }
    \label{fig:Np_results}
  \end{figure*}

  Figure \ref{fig:Np_results} shows the dynamics of the net production $N_{\ce{P}}(t)$.
  Panels~(b) and (c) show the high- ($\Omega = 10 \pi$) and low-frequency ($\Omega = 0.2 \pi$) results, respectively.
  Blue curves correspond to $\gammaAntiClockwise$, while black curves correspond to $\gammaClockwise$ in \cref{fig:driving_sequences}.
  As for the short-time dynamics, a closer look reveals three abrupt changes within each period, which correspond to the switching events of the driving generators as shown in Panels ~(b) and ~(c).
  At stroboscopic times $t=t_0+MT$, the net production $N_{\ce{P}}(t)$ grows approximately linearly after a sufficiently long time as
  \begin{align}
    N_{\ce{P}}(t_0+MT)
    \simeq
    J_{\infty}(\Omega) \cdot MT
    +
    \text{oscillatory terms}.
    \label{eq:N_P_stroboscopic_growth}
  \end{align}
  The coefficient of this linear growth, namely the slope of the envelope, can therefore be interpreted as the asymptotic value of the net current $J_{\infty}(\Omega)$.
  In numerical simulations, however, extracting this slope directly from the definition \cref{eq:J_t} can be inefficient when the initial value $N_{\ce{P}}(t_0)$ leads to a long transient regime.
  We therefore adopt the one-period increment to numerically compute the long-time current
  \begin{align}
      J_{\infty}^{\mathrm{num}}(\Omega) &= \lim_{t\to\infty} J^{\mathrm{num}}(\Omega;t), \\
      J^{\mathrm{num}}(\Omega;t) &= \frac{N_{\ce{P}}(t+T) - N_{\ce{P}}(t)}{T},
      \label{eq:J_for_numerical_calc}
  \end{align}
  and compare its stationary value at large $t$ with the analytic expression for $J_{\infty}(\Omega)$.

  When the driving sequence is reversed from $\gammaAntiClockwise$ to $\gammaClockwise$, the tendency of $N_{\ce{P}}(t)$ is reversed as well.
  The two protocols use the same generators for the same step duration and differ only in their cyclic ordering.
  Therefore, the direction of the current can be controlled by the ordering of the driving sequence.

\begin{figure}[!htbp]
    \centering
    \includegraphics[width=\columnwidth]{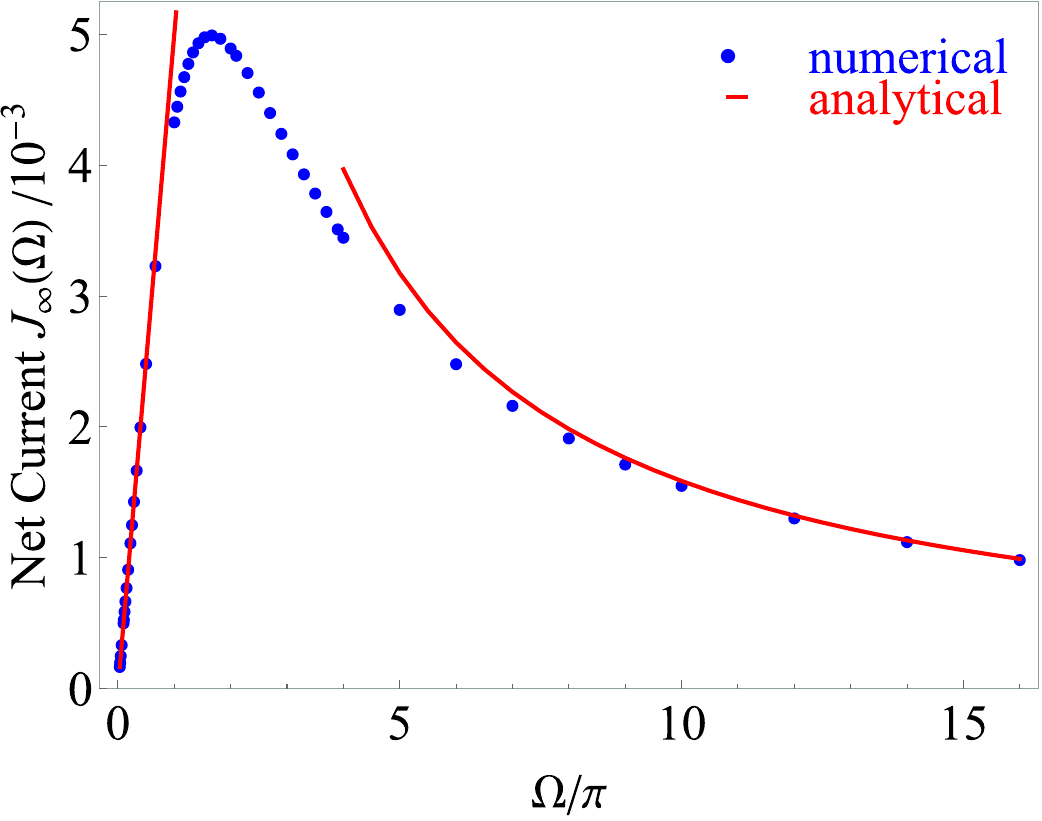}
    \caption{
      Asymptotic net current $J_{\infty}(\Omega)$.
      Blue dots show the numerical results obtained from the steady values of \cref{eq:J_for_numerical_calc}.
      Red curves denote the analytical asymptotics in the high- and low-frequency regimes, \cref{eq:J_geom_low_concrete,eq:J_high_freq}.
      The high- and low-frequency regimes are $\Omega/\pi \gg 1$ and $\Omega/\pi \lesssim 1$, respectively.
    }
    \label{fig:J_Omega_dependence}
  \end{figure}

  Figure \ref{fig:J_Omega_dependence} shows the asymptotic net current $J_{\infty}(\Omega)$ for different driving frequencies $\Omega$.
  The numerical results are shown as blue dots, while the analytical predictions derived in the next two subsections are plotted as red curves.
  In the high-frequency regime, it exhibits an inverse-$\Omega$ decay.
  In the low-frequency regime, $J_{\infty}(\Omega)$ increases linearly with $\Omega$.
  Convergence of $J^{\mathrm{num}}(\Omega;t)$ to a steady value is confirmed for each point by checking the time evolution of $J^{\mathrm{num}}(\Omega;t)$, as shown in \cref{app:numerical_convergene_of_geometric_current}.
  These two asymptotic behaviors are consistent with the high- and low-frequency trends reported previously for continuously driven systems~\cite{Astumian1992-qd,Astumian2001-je,Sinitsyn2007-qo,Ohkubo2008-za,Ohkubo2008-xl,Sinitsyn2009-jy,Takahashi2020-nz}.
  This comparison suggests that similar asymptotic structures persist even under strong, discretely switched driving.

  We now specialize the general expressions of \cref{sec:floquet_theory_counting_statistics} to the $N$-step driving sequence $\gamma = \{\Mstep{0}, \Mstep{1}, \ldots, \Mstep{N-1}\}$.
  We first keep $N$ arbitrary and derive formulas for a general uniform-step discrete protocol, and then substitute the three-step protocol in \cref{eq:rate_constant0,eq:rate_constant1,eq:rate_constant2} when comparing with the numerical results.
  After deriving the uniform-step result, we also remark on the nonuniform-step case where the durations of the steps are not equal.
  We set the initial time to zero, $t_0=0$, and consider stroboscopic times $t=MT$ with a large integer $M$.
  Although the short-time transient dynamics depends on the initial time $t_0$~\cite{Goldman2014-ek}, the long-time averaged current depends only on the driving sequence $\gamma$ and is independent of $t_0$.

\subsection{High-Frequency Regime}
\label{sec:high_freq_regime}

  In the high-frequency regime, the current is governed by the effective Floquet generator obtained from the van Vleck high-frequency expansion:
  \begin{align}
    \Markov^{\mathrm{eff,vV}}(\chi)
    &=
    \Markov_0(\chi)
    +
    \imaginaryi
    \sum_{m \neq 0}
    \frac{
      [\Markov_{-m}(\chi), \Markov_{m}(\chi)]
    }{2m\Omega}
    +
    \mathcal{O}(\Omega^{-2}),
    \label{eq:Markov_eff_vV}
    \\
    \KickMarkov^{\mathrm{vV}}(\chi;t)
    &=
    \imaginaryi
    \sum_{m \neq 0}
    \frac{\Markov_m(\chi)}{m\Omega}
    \napier^{-\imaginaryi m \Omega t}
    +
    \mathcal{O}(\Omega^{-2}),
    \label{eq:KickMarkov_vV}
  \end{align}
  where $\Markov_m(\chi)$ is the $m$-th Fourier component of $\Markov(\chi;t)$.
  The derivation and the general expression for the current are given in \cref{sec:floquet_theory_counting_statistics}.
  For the uniform $N$-step discrete driving,
  \cref{eq:Markov_eff_vV} is transformed into
  \begin{align}
    \displaystyle
      &\Markov^{\mathrm{eff},\mathrm{vV}} (\chi)
      =
      \nonumber \\ & \quad
      \Markov_0
      +
      \imaginaryi
      \sum_{a=1}^{N-1}
      \frac{a^2}{N^3}
      \zeta\left(3, \frac{a}{N}\right)
      \frac{\left[\Markov_{-a}, \Markov_{a}\right]}{\Omega}
      + \mathcal{O}(\Omega^{-2}),
      \label{eq:W_eff_Nstep_1st_order}
    \end{align}
    where $\zeta(s,a)$ is the Hurwitz zeta function.
    See \cref{app:vVE_underUDD} for detailed derivation and the expression for the $\Omega^{-2}$ term.
    In particular, for the three-step protocol, this reduces to
    \begin{align}
      \Markov^{\mathrm{eff},\mathrm{vV}} (\chi)
      =
      \Markov_0
      +
      \imaginaryi\frac{4\pi^3}{81\sqrt{3}}
      \frac{\left[\Markov_{-1}, \Markov_{1}\right]}{\Omega}
      +
      \mathcal{O}(\Omega^{-2}).
      \label{eq:W_eff_3step_1st_order}
    \end{align}

\begin{figure}[!htbp]
    \centering
    \includegraphics[width=\columnwidth]{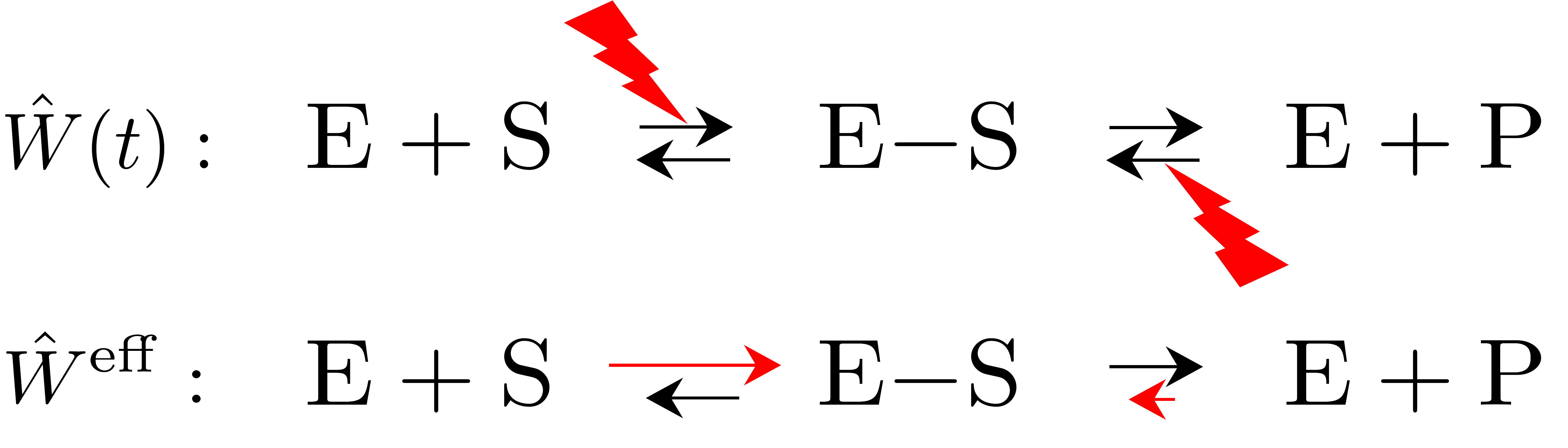}
    \caption{
      Schematic interpretation of the driven system as the effective chemical reactions.
      (Top) Enzymatic reaction with periodically driven reaction rates.
      (Bottom) Effective enzymatic reaction with renormalized reaction rates.
    }
    \label{fig:ER_eff_concept}
  \end{figure}

  Let us consider the situation with \cref{eq:rate_constant0,eq:rate_constant1,eq:rate_constant2} under uniform step duration $\tau=T/3$.
  The explicit form of the effective Floquet generator \cref{eq:W_eff_3step_1st_order} becomes
  \begin{align}
    \Markov^{\mathrm{eff,vV}}(\chi)
    &=
    \Markov_0(\chi)
    +
    \frac{2\pi}{\Omega}
    \begin{pmatrix}
      0 & 0 \\
      \frac{1}{54}(1-\napier^{-\imaginaryi\chi})  & 0
    \end{pmatrix}
    +
    \mathcal{O}(\Omega^{-2}),
    \\
    \Markov_0(\chi)
    &=
    \begin{pmatrix}
      -\frac{8}{3} & 1+\napier^{\imaginaryi\chi} \\
      \frac{4}{3}(1+\napier^{-\imaginaryi\chi}) & -2
    \end{pmatrix},
  \end{align}
  where we consider the protocol $\gammaAntiClockwise$.
  This effective Floquet generator can be interpreted as the following chemical reactions with the effective chemical reaction rates:
  \begin{align}
      \displaystyle
      &\schemestart
        \ce{E} \+ \ce{S}
        \arrow{<=>[$k_1^{\mathrm{eff}}$][$k_{-1}^{\mathrm{eff}}$]}
        \ce{E-S}
        \arrow{<=>[$k_2^{\mathrm{eff}}\napier^{\imaginaryi\chi}$][$k_{-2}^{\mathrm{eff}}\napier^{-\imaginaryi\chi}$]}
        \ce{E} \+ \ce{P},
      \schemestop
      \\
      \shortintertext{where the effective reaction rates are}
      &\qquad
      k_1^{\mathrm{eff}} = \frac{4}{3} + \frac{2\pi}{\Omega}\frac{1}{54}, \quad
      k_2^{\mathrm{eff}} = 1, \\
      &\qquad
      k_{-1}^{\mathrm{eff}} = 1, \quad
      k_{-2}^{\mathrm{eff}} = \frac{4}{3}-\frac{2\pi}{\Omega}\frac{1}{54}.
  \end{align}
  \cref{fig:ER_eff_concept} illustrates the core consequence of this effective description.

  The current operators and the zero-mode of $\Markov^{\mathrm{eff,vV}}(0)$ are
  \begin{align}
    \hat{J}^{\mathrm{eff,vV}}(\chi)
        &=
        \hat{J}_0(\chi)
        +
        \frac{2\pi}{\Omega}
        \begin{pmatrix}
          0 & 0 \\
          \frac{1}{54}\napier^{-\imaginaryi\chi} & 0
        \end{pmatrix}, \\
    \hat{J}_0(\chi)
        &=
        \begin{pmatrix}
          0 & \napier^{\imaginaryi\chi} \\
          -\frac{4}{3}\napier^{-\imaginaryi\chi} & 0
        \end{pmatrix}, \\
    \ket|\psi_0>
        &=
        \begin{pmatrix}
          3/7 \\
          4/7
        \end{pmatrix}.
  \end{align}
  The dynamical contribution vanishes as
  \begin{align}
    \braket<S|\hat{J}_0|\psi_0>=0.
  \end{align}
  Therefore, we obtain the asymptotic current as
  \begin{align}
    J_{\infty}^{(\gammaAntiClockwise)}(\Omega)
    &=
    \frac{1}{126}\frac{2\pi}{\Omega}
    +
    \mathcal{O}(\Omega^{-2}),
    \label{eq:J_high_freq} \\
    J_{\infty}^{(\gammaClockwise)}(\Omega)
    &=
    -\frac{1}{126}\frac{2\pi}{\Omega}
    +
    \mathcal{O}(\Omega^{-2}).
  \end{align}
  Thus, in the high-frequency regime, the asymptotic current exhibits an inverse-$\Omega$ decay and changes sign under reversal of the driving sequence.

\subsection{Low-Frequency Regime}
\label{sec:low_freq_regime}

  In the low-frequency regime, the duration $\tau=T/N$ of each step is sufficiently long compared with the relaxation time of the dynamics generated by $\Mstep{a}$.
  Then the state relaxes to the dominant mode within each interval, and the general low-frequency result \cref{eq:J_low} reduces to
  \begin{align}
    J_{\infty} &= J_{\mathrm{geom}}(\Omega) + J_{\mathrm{dyn}} \label{eq:J_low_3step}, \\
    J_{\mathrm{geom}}(\Omega)
    &=
    \frac{\Omega}{2\pi}\,
    \partial_\chi\phi(\chi)\big|_{\chi=0},
    \label{eq:J_geom_low}
    \\
    \phi(\chi)
    &=
    \ImaginaryPartOf\log
    \prod_{a=0}^{N-1}
    \braket*<\tilde{u}_0^{a+1}(\chi)|u_0^{a}(\chi)>,
    \label{eq:geometric_phase}
    \\
    J_{\mathrm{dyn}}
    &=
    \frac{1}{N}
    \sum_{a=0}^{N-1}
    \braket*<\tilde{u}^a_0|
      \left( -\imaginaryi\partial_\chi \Mstep{a} \right)
    |u^a_0>
    \Big|_{\chi=0}.
    \label{eq:J_dyn_low}
  \end{align}
  Here, $\ket|u_0^a(\chi)>$ and $\bra*<\tilde{u}_0^a(\chi)|$ are the right and left dominant eigenvectors of the generator $\Mstep{a}(\chi)$, and the step index is understood to be cyclic, i.e., $\ket|u_0^N(\chi)> \equiv \ket|u_0^0(\chi)>$.
  We impose the normalization condition
  \begin{align}
    \braket*<\tilde{u}_0^a(\chi)|u_0^a(\chi)>=1.
  \end{align}
  The phase $\phi(\chi)$ is the geometric phase accumulated over a single driving period, which is a classical analogue of the Aharonov--Anandan phase, in the present discretely driven setting.
  The derivative
  \(
    \partial_{\chi}\log
    \prod_{a=0}^{N-1}
    \braket*<\tilde{u}_0^{a+1}|u_0^a>\big|_{\chi=0}
  \)
  is purely imaginary, as shown in \cref{app:pure_imaginary_nature_of_partial_chi_log_Gamma}, so $J_{\mathrm{geom}}(\Omega)$ is real.
  The dynamical contribution $J_{\mathrm{dyn}}$ is also real, because $-\imaginaryi\partial_{\chi}\Mstep{a}(\chi)|_{\chi=0}$ is a real matrix and the zero modes of $\Mstep{a}(0)$ can be chosen real.

  In the low-frequency regime,
  the geometric current is proportional to the driving frequency $\Omega$, because the same geometric phase is accumulated once per period.
  The dynamical current does not depend on $\Omega$.
  This $\Omega$-scaling follows directly from \cref{eq:J_geom_low,eq:J_dyn_low}.

  A significant property of the geometric current is its robustness.
  For example, it does not directly depend on the duration.
  Suppose that the same cyclic sequence $\gamma=\{\Mstep{0},\Mstep{1},\ldots,\Mstep{N-1}\}$ is used, but the duration of step $a$ is $\Delta t_a$ with $\sum_{a=0}^{N-1}\Delta t_a=T$.
  Provided that each duration is still long enough for relaxation to the corresponding dominant mode, the geometric current \cref{eq:J_geom_low} remains unchanged since the geometric phase in \cref{eq:geometric_phase} does not depend on $\Delta t_a$.
  This is in contrast with the dynamical current which is modified as:
  \begin{align}
    J_{\mathrm{dyn}}
    &=
    \sum_{a=0}^{N-1}
    \left(
      -\imaginaryi
      \partial_{\chi}
      \lambda_0^a(\chi)
      \Big|_{\chi=0}
    \right)
    \frac{\Delta t_a}{T},
    \label{eq:J_dyn_low_nonuniform_steps}
  \end{align}
  where $\lambda_0^a(\chi)$ is the dominant eigenvalue of $\Mstep{a}(\chi)$.
  Thus, for a fixed total period $T$ and a fixed cyclic sequence $\gamma$, the geometric contribution is insensitive to redistributing the durations among the steps, whereas the dynamical contribution changes through the weights $\Delta t_a/T$.

  For the three-step protocol in \cref{eq:rate_constant0,eq:rate_constant1,eq:rate_constant2}, the contributions from the three steps are explicitly calculated as
  \begin{align}
    J_{\mathrm{dyn}}^{(\gammaAntiClockwise)}=0,
    \qquad
    J_{\mathrm{geom}}^{(\gammaAntiClockwise)}(\Omega) = \frac{1}{100} \frac{\Omega}{2\pi}
    \propto
    \Omega.
    \label{eq:J_geom_low_concrete}
  \end{align}
  Reversing the driving sequence changes the sign of the geometric contribution:
  \begin{align}
    J_{\mathrm{geom}}^{(\gammaClockwise)}(\Omega)
    =
    -\frac{1}{100}\frac{\Omega}{2\pi}.
  \end{align}
  Thus, in the low-frequency regime, the asymptotic current is linear in $\Omega$ and the sign of the geometric contribution is controlled by the orientation of the driving sequence.

\subsection{Validity of Analytical Results in Two Frequency Regimes}
\label{sec:validity_of_analytical_results}
  The validity of the high-frequency result follows from the truncation of the van Vleck expansion.
  Since the expansion is truncated at order $\Omega^{-1}$, the leading omitted terms are of order $\Omega^{-2}$.
  In the present model, the matrix elements of $\Mstep{a}$ are of order $k_r^a=\mathcal{O}(1)$.
  Thus the expansion is reliable when $\Omega/\pi\gg1$.
  The residual terms introduced in \cref{eq:N_P_low_freq_formal} are sublinear and oscillatory in time, and therefore do not contribute to the long-time current; see \cref{app:rigorous_derivation_of_J_high} for the concrete expression.
  As a result, the leading error in the high-frequency regime is governed by the omitted $\mathcal{O}(\Omega^{-2})$ terms.
  This is also consistent with \cref{fig:J_Omega_dependence}, where the numerical results approach the analytical asymptote as $\Omega$ becomes large.

  The validity of the low-frequency result can be understood as follows (as in Ref.~\cite{Sinitsyn2007-qo}).
  In \cref{sec:floquet_theory_counting_statistics}, we made an approximation in the low-frequency regime by projecting onto the instantaneous dominant mode.
  For the stepwise protocol considered here, this approximation is valid when the duration $\tau=T/N$ of each step is sufficiently long compared with the relaxation time of the dynamics generated by $\Mstep{a}$.
  More precisely, since $\lambda_0^a(0)=0$ and $\RealPartOf{\lambda_1^a(0)}<0$, the nonzero-mode contribution is suppressed relative to the zero-mode contribution by a factor of order
  \begin{align}
    \exp\!\left(
      -|\RealPartOf{(\lambda_1^a(0)-\lambda_0^a(0))}|\tau
    \right).
  \end{align}
  Hence the low-frequency expression is justified when $|\RealPartOf{(\lambda_1^a-\lambda_0^a)}|\tau \gtrsim 1$ for every step $a$.
  For the uniform-step protocol, using $\tau=2\pi/(N\Omega)$, this condition is written as
  \begin{align}
    \Omega
    \lesssim
    \frac{2\pi}{N}
    \min_a
    |\RealPartOf{(\lambda_1^a-\lambda_0^a)}|.
  \end{align}
  With non-uniform durations, the corresponding condition is
  \begin{align}
    |\RealPartOf{(\lambda_1^a-\lambda_0^a)}|\,\Delta t_a
    \gtrsim
    1
  \end{align}
  for every step $a$.
  In the three-step protocol in \cref{eq:rate_constant0,eq:rate_constant1,eq:rate_constant2}, the relevant relaxation scale is of order unity.
  Therefore, the low-frequency approximation is expected to work in the region $\Omega/\pi \lesssim 1$, which is consistent with \cref{fig:J_Omega_dependence}.

%% file: sections/5-conclusion.tex

\section{Conclusion} \label{sec:conclusion}

We considered a Floquet framework for periodically driven continuous-time Markov processes using a counting field.
As a prototypical problem, we focused on the Michaelis--Menten-type chemical reaction process and studied the first cumulant $N_{\ce{P}}(t)$ of the counting variable $n$ and the corresponding long-time current $J_{\infty}(\Omega)$.
We obtained the formal expressions for $N_{\ce{P}}(t)$ and $J_{\infty}(\Omega)$ for arbitrary driving frequency regimes using only the time periodicity of the Markov generator.
We further derived explicit analytical expressions for $J_{\infty}(\Omega)$ in two asymptotic regimes: in the high-frequency regime, where the current is governed by the van Vleck effective Floquet generator, and in the low-frequency regime, where it is described by the instantaneous zero eigenvector of the generator at each time.

We then applied this framework to discretely driven continuous-time Markov processes, motivated by the cAMP production system.
The cAMP production system plays important roles in cellular signaling, and the catalytic state of AC can be viewed as discrete switching among different activity states.
By modeling this situation as a discretely driven Michaelis--Menten-type reaction, we evaluated the current induced by periodic switching of the reaction rates.
In the high-frequency regime, the periodically driven chemical reaction system can be interpreted as an effective chemical reaction system with renormalized reaction rates.
We confirmed that our analytical expressions show good agreement with the numerical results.

Recent advances in optogenetics have made it possible to manipulate light-responsive proteins, including GPCRs in the cAMP production system, with optical inputs.
These developments suggest that the biochemical situation considered in our model may be experimentally realizable using such techniques, where proteins often switch among a discrete set of conformational or activation states.
Because of this intrinsic discreteness, it is important to consider periodically driven dynamics in which the driving itself is discrete.
As a future direction, collaboration with experimentalists will be essential for establishing a more quantitative comparison between the present theoretical framework and experimentally realized biochemical reaction systems.

%% file: sections/10-appendix.tex
\newpage
\appendix
\appendixpage
\addappheadtotoc

\section{Numerical Convergence of Long-Time Current} \label{app:numerical_convergene_of_geometric_current}
\begin{figure*}[htbp]
  \centering
  \begin{minipage}[t]{\columnwidth}
    \centering
    \includegraphics[width=0.9\linewidth]{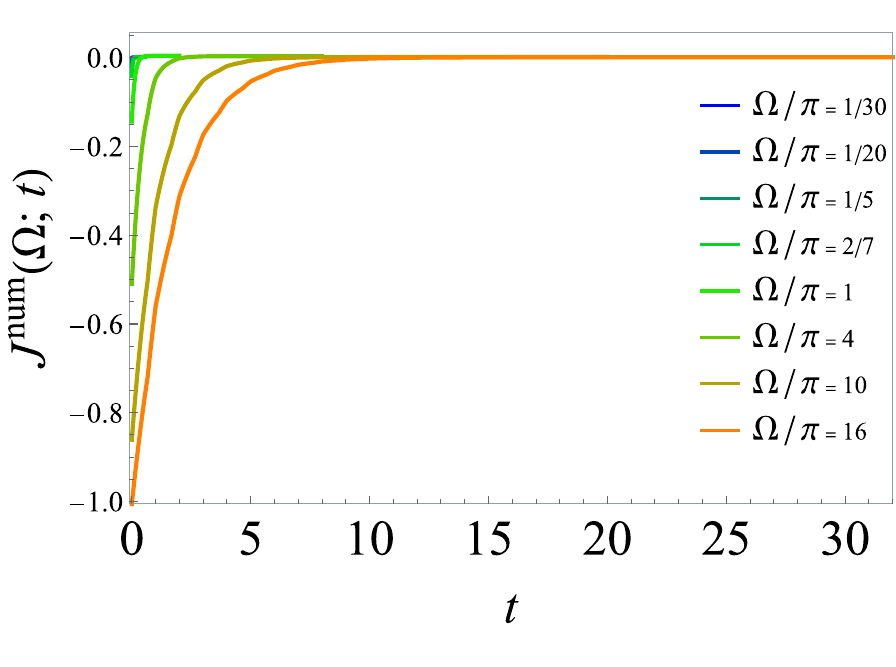}
  \end{minipage}
  \begin{minipage}[t]{\columnwidth}
    \centering
    \includegraphics[width=0.9\linewidth]{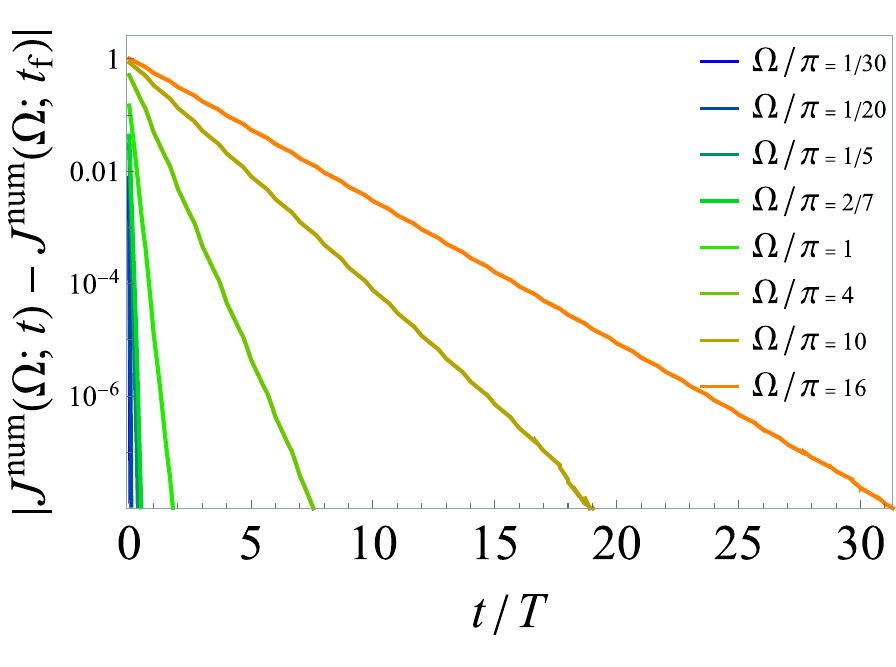}
  \end{minipage}
  \caption{
    Convergence of $J^{\mathrm{num}}(\Omega;t)$ to the long-time current $J_{\infty}(\Omega)$.
    (Left) $J^{\mathrm{num}}(\Omega;t)$ on a linear scale.
    (Right) The difference $|J^{\mathrm{num}}(\Omega;t) - J^{\mathrm{num}}(\Omega; t_{\mathrm{f}})|$ on a logarithmic scale, illustrating the exponential convergence.
    The horizontal axis indicates the normalized time $t / T$.
    Each plot is shown up to the first time $t_\varepsilon$ satisfying $|J^{\mathrm{num}}(\Omega; t)-J^{\mathrm{num}}(\Omega; t_{\mathrm{f}})|\le \varepsilon$, where $t_{\mathrm{f}} $ is the final time of the simulation (here, $t_{\mathrm{f}} = 60T$) and $\varepsilon = 10^{-8}$.
    The converged value $J^{\mathrm{num}}(\Omega; t_\varepsilon)$ defines the long-time current $J_{\infty}(\Omega)$.
  }
  \label{fig:numerical_convergence_of_geometric_current}
\end{figure*}

  Figure~\ref{fig:numerical_convergence_of_geometric_current} shows the time evolution of the current $J^{\mathrm{num}}(\Omega; t)$ defined in \cref{eq:J_for_numerical_calc} from low to high frequencies.
  For visibility, we select frequencies $\Omega/\pi = 1/30, 1/20, 1/5, 2/7, 1, 4, 10, 16$.
  $J^{\mathrm{num}}(\Omega; t)$ exhibits exponential convergence to a steady value as time $t$ increases.
  We determine the time $t_\varepsilon$ at which $J^{\mathrm{num}}(\Omega; t)$ converges to a steady value within a tolerance $\varepsilon$, and we take $J^{\mathrm{num}}(\Omega; t_\varepsilon)$ as the long-time current $J_{\infty}(\Omega)$.

\section{Proof that  \texorpdfstring{$\partial_{\chi}\log \Gamma(\chi)\big|_{\chi=0}$}{partial\_chi\_log\_Gamma(chi)|\_{chi=0}}  is purely imaginary} \label{app:pure_imaginary_nature_of_partial_chi_log_Gamma}

  In this section, we show that $\partial_{\chi}\log \Gamma(\chi)\big|_{\chi=0}$ is purely imaginary, which implies that the geometric current $J_{\mathrm{geom}}(\Omega)$ is real.
  For simplicity, we define
  \begin{align}
    \Gamma(\chi) \equiv \prod_{j=1}^{L} \braket*<\tilde{u}_0(\chi;t_j)|u_0(\chi;t_{j-1})>.
  \end{align}

  Because the counting field enters the generator only through factors of $\napier^{\pm\imaginaryi\chi}$, we have
  \begin{align}
    \Markov (\chi;t)=\Markov (-\chi;t)^{*}.
  \end{align}
  The right eigenvector $\ket*|u_0(\chi;t)>$ of $\Markov(\chi;t)$ satisfies
  \begin{align}
    \Markov(\chi;t)\ket*|u_0(\chi;t)>=\lambda_0(\chi;t)\ket*|u_0(\chi;t)>,
  \end{align}
  where $\lambda_0(\chi;t)$ is the eigenvalue continuously connected to zero at $\chi=0$.
  Complex conjugation gives
  \begin{align}
    \Markov(-\chi;t)\ket*|u_0(\chi;t)>^*=\lambda_0(\chi;t)^*\ket*|u_0(\chi;t)>^*,
  \end{align}
  and we can choose the eigenvectors so that
  \begin{align}
    \ket*|u_0(\chi;t)>=\ket*|u_0(-\chi;t)>^*,
    \qquad
    \bra*<\tilde u_0(\chi;t)|=\bra*<\tilde u_0(-\chi;t)|^*.
  \end{align}
  Therefore, each overlap satisfies
  \begin{align}
    \braket*<\tilde u_0(\chi;t_j)|u_0(\chi;t_{j-1})>
    =
    \braket*<\tilde u_0(-\chi;t_j)|u_0(-\chi;t_{j-1})>^{*},
  \end{align}
  and hence
  \begin{align}
    \Gamma(\chi)=\Gamma(-\chi)^{*}.
  \end{align}
  Choosing the branch of the logarithm that is analytic around $\chi=0$, we have
  \begin{align}
    \log \Gamma(\chi)=\bigl(\log \Gamma(-\chi)\bigr)^{*}.
  \end{align}
  Differentiating both sides with respect to $\chi$ and setting $\chi=0$ yields
  \begin{align}
    \partial_\chi \log \Gamma(\chi)\Big|_{\chi=0}
    =
    -\left(
      \partial_\chi \log \Gamma(\chi)\Big|_{\chi=0}
    \right)^{*}.
  \end{align}
  Therefore, $\partial_\chi \log \Gamma(\chi)\big|_{\chi=0}$ is purely imaginary.

\section{High-Frequency Expansion} \label{app:high_freq_expansion}
  We briefly summarize the high-frequency expansion,
  which is a perturbative expansion in powers of $\Omega^{-1}$.
  Within Floquet theory, the time-evolution operator $\TimeEvolutionOp(\chi; t, t_0)$ can be decomposed as
  \begin{align}
    \TimeEvolutionOp(\chi;t,t_{0})
    &= \hat{V}(\chi;t)\,\exp\!\big(-\imaginaryi\SchH^{\mathrm{eff}}(\chi)(t-t_{0})\big)\,\hat{V}^{-1}(\chi;t_{0}), \\
    \hat{V}(\chi;t)
    &=
      \exp\!\big(-\imaginaryi\KickSch(\chi;t)\big),
      \label{eq:U_highfreq}
  \end{align}
  where $\SchH^{\mathrm{eff}}(\chi)$ is the effective Hamiltonian and $\hat{V}(\chi;t)$ is the micromotion operator~\cite{Shavitt1980-tx,Eckardt2015-qn,Mikami2016-yv,Goldman2014-ek}.
  The explicit forms of the kick operator and the effective Hamiltonian up to the first order in $\Omega^{-1}$ are given by~\cite{Kitagawa2011-ia}
  \begin{align}
    \KickSch(\chi; t)
    &=
      \KickSch_0
      + \imaginaryi \sum_{m \neq 0}
        \frac{\SchH_m}{m\Omega}
        \napier^{-\imaginaryi m\Omega t}
      + \mathcal{O}(\Omega^{-2}),
    \label{eq:Kick_op_vV} \\
    \SchH^{\mathrm{eff}}(\chi)
    &=
      \SchH_0
      +
      \sum_{m \neq 0}
        \frac{
          \left[\SchH_{-m}, \SchH_m\right]
        }{2m\Omega}
      + \left[\imaginaryi\KickSch_0, \SchH_0\right]
    \nonumber \\
    & \quad + \mathcal{O}(\Omega^{-2}). \label{eq:definition_H_eff_vV}
  \end{align}

  In order to derive the high-frequency expansion of the effective Hamiltonian and the kick operator, the general logarithmic-derivative identity for the exponential operators is used.
  The identity was originally derived for $\Omega(t)=\log U(t)$ by Magnus~\cite{Magnus1954-rh} and has been reformulated in a modern operator-theoretic form in Refs.~\cite{Casas2001-ni,Blanes2009-zn,Bauer2013-hi,Kitagawa2011-ia,Eckardt2015-qn,Bukov2015-ux,Ebrahimi-Fard2025-uk}:
  \begin{align}
    \frac{\partial}{\partial t} \KickSch(t)
    =
    \sum_{m=0}^{\infty}
      \frac{B_m}{m!}
      (- \imaginaryi \mathrm{ad}_{\KickSch(t)})^m
      \left[
        \SchH(t) - (-1)^{m} \SchH^{\mathrm{eff}}
      \right],
  \end{align}
  where $B_m$ is the $m$-th Bernoulli number and $\mathrm{ad}_{\hat{X}}$ is defined as $\mathrm{ad}_{\hat{X}} \hat{Y} \equiv [\hat{X}, \hat{Y}]$.
  The next step is to expand $\SchH^{\mathrm{eff}}$ and $\KickSch(t)$ in terms of $\Omega^{-1}$ as
  \begin{align}
    \SchH^{\mathrm{eff}}
    &=
      \SchH^{\mathrm{eff},(0)}
      +
      \SchH^{\mathrm{eff},(1)}
      +
      \mathcal{O}(\Omega^{-2}), \\
    \KickSch(t)
    &=
      \sum_{m \in \mathbb{Z}} \KickSch_m
      \napier^{-\imaginaryi m \Omega t},
    \quad
    \KickSch_m
    =
      \sum_{p=0}^{\infty}
      \KickSch_m^{(p)},
  \end{align}
  where the superscript $(p)$ denotes the order in $\Omega^{-1}$.
  By comparing the coefficients for the first order of $\Omega^{-1}$ in the $m$-th Fourier component, we get
  \begin{align}
    - \imaginaryi m \Omega \KickSch_m^{(1)}
    =
    B_0 \left[
      \SchH_m - \delta_{m,0} \SchH^{\mathrm{eff},(0)}
    \right],
  \end{align}
  and thus,
  \begin{align}
    \KickSch_m^{(1)} &= \imaginaryi \frac{\SchH_m}{m\Omega} \quad (m \neq 0), \quad \SchH^{\mathrm{eff},(0)} = \SchH_0.
  \end{align}
  At the next order in $\Omega^{-1}$, the zeroth Fourier component gives
  \begin{align}
    \SchH^{\mathrm{eff},(1)}
    =
      \sum_{m \neq 0}
        \frac{[\SchH_{-m}, \SchH_{m}]}{2m\Omega}
      +
      \left[\imaginaryi\KickSch_0^{(1)}, \SchH_0\right].
  \end{align}

  We denote $\Markov^{\mathrm{eff}}(\chi) \equiv -\imaginaryi \SchH^{\mathrm{eff}}(\chi)$ and $\KickMarkov(\chi; t) \equiv -\imaginaryi \KickSch(\chi; t)$ to relate the notation in the quantum case to that in the stochastic process case.
  Based on the above results,
  the explicit forms of $\Markov^{\mathrm{eff}}(\chi)$ and $\KickMarkov(\chi; t)$ up to the first order in $\Omega^{-1}$ are
  \begin{align}
    \Markov^{\mathrm{eff}}(\chi)
    &=
      \Markov_0
      + \imaginaryi
      \sum_{m \neq 0}
        \frac{[\Markov_{-m}, \Markov_{m}]}{2m\Omega}
      - \left[\KickMarkov^{(1)}_0, \Markov_0\right]
      \nonumber \\
    &\quad
      + \mathcal{O}(\Omega^{-2}),
    \label{eq:H_eff_vV_explicit} \\
    \KickMarkov(\chi; t)
    &=
      \KickMarkov_0
      +\imaginaryi
      \sum_{m \neq 0}
        \frac{\Markov_m}{m\Omega}
        \napier^{-\imaginaryi m\Omega t}
      + \mathcal{O}(\Omega^{-2}).
    \label{eq:Kick_op_vV_explicit}
  \end{align}
  The remaining higher-order terms can be derived in the same manner.
  For the van Vleck high-frequency expansion, we set $\KickMarkov_0 = 0$, and the following explicit forms (\cref{eq:Markov_eff_vV,eq:KickMarkov_vV}) are obtained:
  \begin{align}
    \Markov^{\mathrm{eff,vV}}(\chi)
    &=
      \Markov_0
      + \imaginaryi\sum_{m \neq 0}
        \frac{[\Markov_{-m}, \Markov_{m}]}{2m\Omega}
      + \mathcal{O}(\Omega^{-2}), \\
    \KickMarkov^{\mathrm{vV}}(\chi; t)
    &=
      \imaginaryi \sum_{m \neq 0}
        \frac{\Markov_m}{m\Omega}
        \napier^{-\imaginaryi m \Omega t}
      + \mathcal{O}(\Omega^{-2}).
  \end{align}

\section{The van Vleck High-Frequency Expansion under the Uniform Discrete Driving} \label{app:vVE_underUDD}
    In this section, we derive the expression for the van Vleck high-frequency expansion under the uniform discrete $N$-step driving.
    We start from the van Vleck expanded form of the effective generator:
    \begin{align}
        \Markov^{\mathrm{eff,vV}} &=
            \Markov_0 + \Markov^{(1)} + \Markov^{(2)} + \mathcal{O}(\Omega^{-3}), \\
        \Markov^{(1)} &= \imaginaryi \sum_{m \neq 0} \frac{[\Markov_{-m}, \Markov_m]}{2m\Omega},
        \label{eq:W_1st_order}
        \\
        \Markov^{(2)} &=
            - 
            \sum_{m \neq 0} \frac{[[\Markov_{-m}, \Markov_0], \Markov_m]}{2 m^2 \Omega^2} \nonumber\\
            & \quad
            - \sum_{m \neq 0} \sum_{n \neq 0, m}
            \frac{[[\Markov_{-m}, \Markov_{m-n}], \Markov_n]}{3mn \Omega^2}.
        \label{eq:W_2nd_order}
    \end{align}
    We show that, under the uniform discrete $N$-step driving, the expression is explicitly rewritten as
    \begin{widetext}
    \begin{align}
      \Markov^{\mathrm{vV}}_{\mathrm{eff}} (\chi)
      &=
      \Markov_0
      +
      \imaginaryi
      \sum_{a=1}^{N-1}
      \frac{a^2}{N^3}
      \zeta\left(3, \frac{a}{N}\right)
      \frac{[\Markov_{-a}, \Markov_{a}]}{\Omega}
      \nonumber
      \\ & \quad
      -
      \sum_{a=1}^{N-1}
      \frac{a^2}{N^4}
      \zeta\left(4, \frac{a}{N}\right)
      \frac{[[\Markov_{-a}, \Markov_0], \Markov_{a}]
        +
        [[\Markov_{a}, \Markov_0], \Markov_{-a}]}{2\Omega^2}
      \nonumber
      \\ & \quad
      -
        \sum_{\substack{a,b=1\\a\neq b}}^{N-1}
        \frac{ab(a-b)}{N^5}
        \mathcal K\left(\frac{a}{N},\frac{b}{N}\right)
        \frac{[[\Markov_{-a},\Markov_{a-b}],\Markov_b]}{3\Omega^2}
      +
      \mathcal{O}(\Omega^{-3}),
      \label{eq:W_eff_Nstep}
      \\
      \shortintertext{where,}
      \zeta(s, z) &= \sum_{m = 0}^\infty \frac{1}{(m + z)^s},
      \quad
      \mathcal K (x, y) = \frac{\pi^5(3\cos(\pi x) \cos(\pi y) + \sin(\pi x) \sin(\pi y))}{3\sin^2(\pi x) \sin^2(\pi y) \sin(\pi (x-y))}.
    \end{align}
    Here, $\zeta(s,z)$ is known as the Hurwitz zeta function.
    \end{widetext}
    The proof is a straightforward, albeit somewhat lengthy, calculation.
    The main ingredient is the following key identity:
    \begin{align}
        \Markov_m &= \begin{cases}
            \displaystyle
            \Markov_0, &(m=0)\\
            0, & (m \equiv 0, \, m \neq 0) \\
            \frac{u}{m}\Markov_u, & (m \not\equiv 0)
        \end{cases}
        \label{eq:key_congruence_relation} \\
        \shortintertext{which follows from}
        \Markov_m
            &= \frac{1}{T} \int_0^T \mathrm{d}t \Mstep{a} \napier^{\imaginaryi m \Omega t}
            \nonumber\\
            &= \frac{1}{T} \sum_{a=1}^N \int_{(a-1)T/N}^{aT/N} \mathrm{d}t \Mstep{a} \napier^{\imaginaryi m \Omega t}
            \nonumber\\
            &=
                \frac{1}{m}
                \frac{\napier^{\imaginaryi m \Omega T/N} - 1}{2\pi \imaginaryi} \sum_{a=1}^N \napier^{\imaginaryi (a-1) m \Omega T/N} \Mstep{a},
    \end{align}
    where $m$ and $u$ are integers with $m \neq 0$ and $u \equiv m \pmod N$.
    This identity follows from the root-of-unity periodicity of the exponential factors.
    First, let us transform $\Markov^{(1)}$ by using \cref{eq:key_congruence_relation} as:
    \begin{align}
        \sum_{m\neq0} \frac{[\Markov_{-m}, \Markov_{m}]}{2m\Omega}
          &=
            \sum_{a=1}^{N-1} \sum_{m > 0, m \equiv a}
            \frac{\left[\Markov_{-m}, \Markov_{m}\right]}{m\Omega}
        \nonumber \\
          &=
            \sum_{a=1}^{N-1}
            \sum_{m > 0, m \equiv a}
            \frac{a^2}{m^3}
            \frac{[\Markov_{-a}, \Markov_a]}{\Omega}
        \nonumber \\
        &=
            \sum_{a=1}^{N-1}
            \frac{[\Markov_{-a}, \Markov_a]}{\Omega}
            a^2
            \sum_{k=0}^\infty \frac{1}{(Nk+a)^3}
        \nonumber \\
        &=
            \sum_{a=1}^{N-1}
            \frac{a^2}{N^3}
            \zeta\left(3,\frac{a}{N}\right)
            \frac{[\Markov_{-a}, \Markov_a]}{\Omega}.
            \label{eq:W_1st_order_result}
    \end{align}
    In the similar manner, we can transform the second order term $\Markov^{(2)}$:
    \begin{align}
        & - 
            \sum_{m \neq 0} \frac{[[\Markov_{-m}, \Markov_0], \Markov_m]}{2 m^2 \Omega^2} \nonumber \\
        &=
            -
            \sum_{a=1}^{N-1}
            \sum_{m>0,m\equiv a}
            \frac{a^2}{m^4}
            \frac{[[\Markov_{-a}, \Markov_0], \Markov_a] + [[\Markov_{a}, \Markov_0], \Markov_{-a}]}{2 \Omega^2}
        \nonumber \\
        &=
            -
            \sum_{a=1}^{N-1}
            \frac{a^2}{N^4}
            \zeta\left(4, \frac{a}{N}\right)
            \frac{[[\Markov_{-a}, \Markov_0], \Markov_a] + [[\Markov_{a}, \Markov_0], \Markov_{-a}]}{2 \Omega^2},
            \label{eq:W_2nd_order_result_1}
    \end{align}
    which is the first term of $\Markov^{(2)}$, and the second term is
    \begin{align}
        &- \sum_{m \neq 0} \sum_{n \neq 0, m}
            \frac{[[\Markov_{-m}, \Markov_{m-n}], \Markov_n]}{3mn \Omega^2}
        \nonumber \\
        &=
            -
            \sum_{\substack{m,n\in \mathbb{Z}\setminus \{0\},\\m-n\not\equiv0}}
            \frac{[[\Markov_{-m}, \Markov_{m-n}], \Markov_n]}{3mn \Omega^2}
            \nonumber \\
            &\quad
                -
                \sum_{\substack{m,n\in \mathbb{Z}\setminus \{0\}\\m-n\equiv0,m\neq n}}
                \frac{[[\Markov_{-m}, \Markov_{m-n}], \Markov_n]}{3mn \Omega^2}
                \nonumber \\
        &=
            -
            \sum_{\substack{a,b=1\\a\neq b}}^{N-1}
            \frac{ab(a-b)}{N^5}
            \mathcal K\left(\frac{a}{N}, \frac{b}{N}\right)
            \frac{[[\Markov_{-a}, \Markov_{a-b}], \Markov_b]}{3\Omega^2}
            \nonumber\\
            &\quad
                +
                \quad
                0,
            \label{eq:W_2nd_order_result_2}
        \\
        \shortintertext{where we denote}
        \mathcal K(x,y)
        &=
            \sum_{\substack{u,v\in\mathbb{Z}}}
            \frac{1}{(u + x)^2 (v + y)^2(u-v+x-y)}.
    \end{align}
    Since $a/N, b/N$ are non-integer, we can utilize the identities
    \begin{align}
        \sum_{k \in \mathbb{Z}} \frac{1}{k + z}
            = \pi \cot(\pi z), \quad
        \sum_{k \in \mathbb{Z}} \frac{1}{(k+z)^2}
            = \pi^2 \csc^2(\pi z),
        \label{eq:useful_identities}
    \end{align}
    at $z = a/N, b/N$.
    After partial fraction decomposition, we obtain
    \begin{align}
        \frac{1}{V^2(U-V)}
        &=
            \frac{1}{U^2V}
            + \frac{1}{UV^2}
            + \frac{1}{U^2(U-V)},
    \end{align}
    where
    \begin{align}
        U = u+x, \quad V = v+y.
    \end{align}
    Then, for fixed $u$,
    \begin{align}
        &\sum_{v \in \mathbb{Z}}
        \frac{1}{V^2(U-V)}
        \nonumber \\
        &=
            \frac{1}{U^2}
            \sum_{v \in \mathbb{Z}}
            \left(
                \frac{1}{V}
                +
                \frac{1}{U-V}
            \right)
            +
            \frac{1}{U}
            \sum_{v \in \mathbb{Z}}
                \frac{1}{V^2}
        \nonumber \\
        &=
            \frac{\pi (\cot(\pi y) + \cot(\pi (x-y)))}{U^2}
            +
            \frac{\pi^2\csc^2(\pi y)}{U}.
    \end{align}
    Substituting this into $\mathcal K(x,y)$, we get
    \begin{align}
        \mathcal{K} (x,y)
        &=
            \pi (\cot(\pi y) + \cot(\pi (x-y)))
            \sum_{u \in \mathbb{Z}} \frac{1}{U^4}
        \nonumber \\
        &\quad
            +
            \pi^2\csc^2(\pi y)
            \sum_{u \in \mathbb{Z}} \frac{1}{U^3}.
    \end{align}
    The remaining sums are obtained by differentiating \eqref{eq:useful_identities}:
    \begin{align}
        \sum_{u \in \mathbb{Z}} \frac{1}{U^3}
        &=
            \pi^3 \cot(\pi x) \csc^2(\pi x),
        \\
        \sum_{u \in \mathbb{Z}} \frac{1}{U^4}
        &=
            \frac{\pi^4}{3} \csc^2(\pi x)
            (1 + 3\cot^2(\pi x)).
            \label{eq:W_2nd_order_result_3}
    \end{align}
    Therefore,
    \begin{align}
        &\mathcal{K}(x,y) \nonumber \\
        &=
            \frac{\pi^5}{3}
            (
                \cot(\pi y)
                +
                \cot(\pi (x-y))
            )
            \csc^2(\pi x)
            (1 + 3\cot^2(\pi x))
        \nonumber \\
        & \quad
            +
            \pi^5
            \cot(\pi x)\csc^2(\pi x)\csc^2(\pi y).
    \end{align}
    Using elementary trigonometric identities, this can be simplified to
    \begin{align}
        \mathcal K (x, y) = \frac{\pi^5(3\cos(\pi x) \cos(\pi y) + \sin(\pi x) \sin(\pi y))}{3\sin^2(\pi x) \sin^2(\pi y) \sin(\pi (x-y))}.
        \label{eq:Kxy}
    \end{align}
    With \cref{eq:W_1st_order_result,eq:W_2nd_order_result_1,eq:W_2nd_order_result_2,eq:W_2nd_order_result_3,eq:Kxy}, we obtain \cref{eq:W_eff_Nstep}.

    For $N=3$ case, \cref{eq:W_eff_Nstep} becomes
    \begin{widetext}
    \begin{align}
        \Markov^{\mathrm{eff,vV}}
          =
          \Markov_0
          +
          \imaginaryi
          \frac{4\pi^3}{81\sqrt{3}}
          \frac{[\Markov_{-1}, \Markov_{1}]}{\Omega} 
          -
          \left(
          \frac{1}{\sqrt{3}}
          \ClausenFn_4
            \left(\frac{2\pi}{3}\right)
          +
          \frac{8\pi^4}{3^7}
          \right)
          \frac{[[\Markov_{-1}, \Markov_0], \Markov_{1}] + [[\Markov_{1}, \Markov_0], \Markov_{-1}]}{\Omega^2}
          +
          \mathcal{O}(\Omega^{-3}),
          \label{eq:W_eff_N_3}
    \end{align}
    which follows immediately by using the relations
    \begin{align}
        \zeta\left(3,\frac{1}{3}\right) &= 13 \zeta (3) + \frac{2\pi^3}{3\sqrt{3}},
        \quad
        \zeta\left(3,\frac{2}{3}\right) = 13 \zeta (3)- \frac{2\pi^3}{3\sqrt{3}}, \nonumber \\
        \zeta\left(4,\frac{1}{3}\right) &= 40\zeta(4) + 27\sqrt{3} \ClausenFn_4\left( \frac{2\pi}{3} \right),
        \quad
        \zeta\left(4,\frac{2}{3}\right) = 40\zeta(4) - 27\sqrt{3} \ClausenFn_4\left( \frac{2\pi}{3} \right),
        \quad
        \ClausenFn_s(x) = \sum_{m=1}^{\infty} \frac{\sin(mx)}{m^s}.
        \nonumber
    \end{align}
    \end{widetext}
    $\ClausenFn_s(x)$ is called the Clausen function.

\section{Derivation of \texorpdfstring{\cref{eq:J_high_freq_general}}{Eq.~\ref*{eq:J_high_freq_general}}} \label{app:rigorous_derivation_of_J_high}
    In this section, we see that the direct calculation of \cref{eq:J_formal} in the high-frequency regime can be simplified as \cref{eq:J_high_freq_general}.
    We start from the identity
    \begin{align}
      -\imaginaryi
      \partial_{\chi}
      \TimeEvolutionOp(\chi;t,t_0)
      =
      \int_{t_0}^{t}
      \mathrm{d}s\,
      \TimeEvolutionOp(\chi;t,s)
      \hat{J}(\chi;s)
      \TimeEvolutionOp(\chi;s,t_0),
      \label{eq:partial_U_general}
    \end{align}
    where $\hat{J}(\chi;t)\equiv-\imaginaryi\partial_\chi\Markov(\chi;t)$.
    This formula follows from \cref{eq:partial_U_general_formula}.

    By applying \cref{eq:partial_U_general} to the generating function \cref{eq:Z} and using the decomposition \cref{eq:U_Floquet_tripartitioned_chi}, we obtain
    \begin{align}
      &N_{\ce{P}}(t)-N_{\ce{P}}(t_0)
      \nonumber\\
      &=
      \braket<S|
        \int_{t_0}^{t}
        \mathrm{d}s\,
        \hat{\mathcal{J}}(0;s)\,
        \napier^{\Markov^{\mathrm{eff,vV}}(0)(s-t_0)}
      |\Psi(0;t_0)>
      ,
      \label{eq:N_P_high_freq_general}
    \end{align}
    where
    \begin{align}
      \hat{\mathcal{J}}(0;s)
      &\equiv
      \hat{V}^{-1}(0;s)
      \hat{J}(0;s)
      \hat{V}(0;s),
      \\
      \ket|\Psi(0;t_0)>
      &\equiv
      \hat{V}^{-1}(0;t_0)
      \ket|\psi(0;t_0)>.
    \end{align}
    Using the Baker--Campbell--Hausdorff formula and
    $\hat{J}(0;s)=\sum_m\hat{J}_m\napier^{-\imaginaryi m\Omega s}$ with
    $\hat{J}_m\equiv-\imaginaryi\partial_\chi\Markov_m(\chi)|_{\chi=0}$, we have
    \begin{align}
      \hat{\mathcal{J}}(0;s)
      &=
      \sum_m
      \hat{J}_m
      \napier^{-\imaginaryi m\Omega s}
      +
      \imaginaryi
      \sum_{l\neq0}
      \sum_m
      \frac{[\Markov_{-l},\hat{J}_m]}{l\Omega}
      \napier^{\imaginaryi(l-m)\Omega s} \nonumber \\
      &\quad +
      \mathcal{O}(\Omega^{-2}).
      \label{eq:Jcal_expanded_general}
    \end{align}
    The dependence on $s$ remains only through the exponential factors, and $\hat{W}_m$ and $\hat{J}_m$ no longer depend on $s$.
    Thus, the long-time behavior of \cref{eq:N_P_high_freq_general} is governed by the integral of the matrix exponential associated with $\Markov^{\mathrm{eff,vV}}(0)$.
    Let us assume the one-period stochastic propagator $\hat{U}(0;T,0)$ is irreducible.
    Since it has positive diagonal entries for a finite-state continuous-time Markov process with finite transition rates, it is primitive.
    The Perron--Frobenius theorem then implies that the stationary mode is simple and that all other modes decay.
    Equivalently, after choosing the logarithm branch connected to $\chi=0$, the exact effective Floquet generator has a simple zero eigenvalue and other eigenvalues have negative real parts.
    The van Vleck generator $\Markov^{\mathrm{eff,vV}}(0)$ is used as the high-frequency approximation to this exact effective Floquet generator within the order considered.
    This is the situation in which the Drazin pseudoinverse~\cite{Drazin1958-on,Horn2012-kv,Ben-Israel2003-ej} separates the zero-mode part from the nonzero-mode part:
    \begin{align}
        &\int_{0}^{\Delta t}
        \mathrm{d}\sigma\,
        \napier^{\Markov^{\mathrm{eff,vV}}(0)\sigma}
        = \nonumber \\
          &\quad
          \Delta t\,\ProjectorP
          +
          \left(
            \napier^{\Markov^{\mathrm{eff,vV}}(0)\Delta t}
            -
            \hat{I}
          \right)
          \DrazinInverse{\Markov^{\mathrm{eff,vV}}(0)},
          \label{eq:Drazin_integral_general} \\
        \shortintertext{and for \(m\neq0\),}
        &\int_{0}^{\Delta t}
        \mathrm{d}\sigma\,
        \napier^{
          \left(
            \Markov^{\mathrm{eff,vV}}(0)
            -
            \imaginaryi m\Omega\hat{I}
          \right)\sigma
        } =
        \nonumber\\
        &\quad
        \frac{1-\napier^{-\imaginaryi m\Omega\Delta t}}{\imaginaryi m\Omega}
        \ProjectorP +
        \left(
          \napier^{\hat{B}_m\Delta t}
          -
          \ProjectorQ
        \right)
        \DrazinInverse{\hat{B}_m}
        \ProjectorQ,
        \label{eq:integral_2_evaluated}
        \\
        &
        \hat{B}_m = \ProjectorQ\Markov^{\mathrm{eff,vV}}(0)\ProjectorQ - \imaginaryi m\Omega \ProjectorQ,
    \end{align}
    where $\ProjectorP$ is the projection operator onto the zero-mode subspace of $\Markov^{\mathrm{eff,vV}}(0)$, and $\ProjectorQ=\hat{I}-\ProjectorP$ is the projection operator onto the corresponding nonzero-mode subspace (see \cref{app:Drazin_pseudoinverse} for details).
    For the oscillatory components with $m\neq0$, the eigenvalues of $\hat{B}_m$ on the nonzero-mode subspace are $w_\alpha-\imaginaryi m\Omega$, where $\operatorname{Re}w_\alpha<0$ in the high-frequency approximation described above.
    Thus, these eigenvalues cannot vanish, and $\hat{B}_m$ is invertible on the nonzero-mode subspace.
    Hence the corresponding terms are bounded, with decaying transients in the nonzero-mode sector.
    Therefore, only the first term on the right-hand side of \cref{eq:Drazin_integral_general} grows linearly in $\Delta t$.
    Using \cref{eq:Jcal_expanded_general,eq:N_P_high_freq_general,eq:Drazin_integral_general}, we extract the term proportional to $\Delta t$.
    This gives
    \begin{align}
      N_{\ce{P}}(t)-N_{\ce{P}}(t_0)
      =
      \Delta t\,J_{\infty}(\Omega)
      +
      \delta N(\Delta t,t_0,\Omega).
      \label{eq:N_P_high_freq_linear_growth}
    \end{align}
    The explicit forms of $J_{\infty}(\Omega)$ and $\delta N(\Delta t,t_0,\Omega)$ are written as
    \begin{align}
      &J_{\infty}(\Omega)
      = \braket<S|\hat{\mathcal{A}}_0\ProjectorP|\Psi(0;t_0)>, \\
      &\delta N(\Delta t,t_0,\Omega)
      = \nonumber \\
      &\quad \braket*<S|
      \left(
        \hat{\mathcal{A}}_0\,\hat{\mathcal{R}}_0(\Delta t)
        +
        \sum_{m\neq0}
        \napier^{-\imaginaryi m\Omega t_0}
        \hat{\mathcal{A}}_m\,\hat{\mathcal{R}}_m(\Delta t)
      \right)
      |\Psi(0;t_0)>,
      \label{eq:delta_N}
    \end{align}
    where
    \begin{align}
      \hat{\mathcal{A}}_m
      &\equiv
      \hat{J}_m
      +
      \imaginaryi
      \sum_{l\neq0}
      \frac{[\Markov_{-l},\hat{J}_{l+m}]}{l\Omega}
      +
      \mathcal{O}(\Omega^{-2}),
      \\
      \hat{\mathcal{R}}_0(\Delta t)
      &\equiv
      \left(
        \napier^{\Markov^{\mathrm{eff,vV}}(0)\Delta t}
        -
        \hat{I}
      \right)
      \DrazinInverse{\Markov^{\mathrm{eff,vV}}(0)},
      \\
      \hat{\mathcal{R}}_m(\Delta t)
      &\equiv
      \frac{\hat{I} - \napier^{-\imaginaryi m\Omega\Delta t}}{\imaginaryi m \Omega} \ProjectorP +
      \left(
          \napier^{\hat{B}_m\Delta t}
          -
          \ProjectorQ
        \right)
        \DrazinInverse{\hat{B}_m}
        \ProjectorQ.
    \end{align}
    Since the relations $\ProjectorP=\ketbra*|\psi_0><S|$ and $\bra<S|\hat{V}^{-1}(0;t_0)=\bra<S|$ hold, we get
    \begin{align}
        \ProjectorP\ket|\Psi(0;t_0)> = \ket|\psi_0>.
    \end{align}
    Using this and $\hat{\mathcal{A}}_0 = \hat{J}^{\mathrm{eff}}$, we finally obtain
    \begin{align}
        &J_{\infty}(\Omega) = \braket*<S|\hat{J}^{\mathrm{eff}}(0)|\psi_0>, \\
        &\delta N(\Delta t, t_0, \Omega) =  \braket*<S|
            \hat{J}^{\mathrm{eff}} \hat{\mathcal{R}}_0(\Delta t)
        |\Psi(0;t_0)> \nonumber \\
        &\quad
        +
        \braket*<S|
            \sum_{m\neq0}
            \hat{\mathcal{A}}_m
            \frac{\napier^{-\imaginaryi m\Omega t_0} - \napier^{-\imaginaryi m\Omega t}}{\imaginaryi m \Omega}
        |\psi_0> \nonumber \\
        &\quad
        +
        \braket*<S|
            \sum_{m\neq0}
            \hat{\mathcal{A}}_m
            \napier^{-\imaginaryi m \Omega t_0}
            \left(
              \napier^{\hat{B}_m\Delta t}
              -
              \ProjectorQ
            \right)
            \DrazinInverse{\hat{B}_m}
            \ProjectorQ
        |\Psi(0;t_0)>.
        \label{eq:delta_N_appendix}
    \end{align}
    In the long-time limit, the nonzero-mode contributions remain bounded, up to decaying transients, while the $m\neq0$ terms oscillate with finite amplitude.
    Therefore, $\delta N(\Delta t,t_0,\Omega)/\Delta t \to 0$ as $\Delta t \to \infty$.

  \section{Derivations of \texorpdfstring{\cref{eq:Drazin_integral_general}}{Eq.~\ref*{eq:Drazin_integral_general}} and \texorpdfstring{\cref{eq:integral_2_evaluated}}{Eq.~\ref*{eq:integral_2_evaluated}}} \label{app:derivations}
    We derive \cref{eq:Drazin_integral_general,eq:integral_2_evaluated} together:
    \begin{align*}
      &\int_{0}^{\Delta t} \mathrm{d} \sigma \,
      \napier^{(\Markov^{\mathrm{eff}, \mathrm{vV}}(0) - \imaginaryi m \Omega) \sigma} \nonumber \\
    &\quad =
        C(\Delta t)\, \ProjectorP + \left(\napier^{\hat{B}_m \Delta t} - \ProjectorQ\right) \DrazinInverse{\hat{B}_m} \ProjectorQ, \\
    &C(\Delta t)
        \equiv
        \begin{cases}
            \Delta t & (\text{if } m = 0), \\
            \frac{1-\napier^{-\imaginaryi m\Omega\Delta t}}{\imaginaryi m\Omega} & (\text{if } m \neq 0).
        \end{cases}
    \end{align*}
    Let
    \begin{align}
        \hat{L}(\Delta t) &\equiv
            \int_{0}^{\Delta t} \mathrm{d} \sigma \,
            \napier^{(\Markov^{\mathrm{eff}, \mathrm{vV}}(0) - \imaginaryi m \Omega) \sigma}, \\
        \hat{R}(\Delta t) &\equiv
            C(\Delta t)\, \ProjectorP + \left(\napier^{\hat{B}_m \Delta t} - \ProjectorQ\right) \DrazinInverse{\hat{B}_m} \ProjectorQ.
    \end{align}
    Showing $\hat{L}(\Delta t) = \hat{R}(\Delta t)$ is sufficient.

    Since $\exp(\Markov \sigma)$ is continuous with respect to $\sigma$, $\hat{L}(\Delta t)$ and $\hat{R}(\Delta t)$ are also continuous with respect to $\Delta t$.
      The derivative of $\hat{L}(\Delta t), \hat{R}(\Delta t)$ with respect to $\Delta t$ are
      \begin{align}
        \frac{\mathrm{d}}{\mathrm{d} (\Delta t)} \hat{L}(\Delta t)
            &=
            \napier^{(\Markov^{\mathrm{eff}, \mathrm{vV}}(0) - \imaginaryi m \Omega) \Delta t} \nonumber \\
            &=
            (\ProjectorP + \ProjectorQ) \napier^{(\Markov^{\mathrm{eff}, \mathrm{vV}}(0) - \imaginaryi m \Omega) \Delta t} (\ProjectorP + \ProjectorQ) \nonumber \\
            &=
            \napier^{-\imaginaryi m \Omega \Delta t}\, \ProjectorP
            +
            \ProjectorQ\, \napier^{(\Markov^{\mathrm{eff}, \mathrm{vV}}(0) - \imaginaryi m \Omega) \Delta t}\, \ProjectorQ \nonumber \\
            &=
            \napier^{-\imaginaryi m \Omega \Delta t}\, \ProjectorP
            +
            \napier^{\hat{B}_m \Delta t} \ProjectorQ, \\
            
        \frac{\mathrm{d}}{\mathrm{d} (\Delta t)} \hat{R}(\Delta t)
            &=
            \napier^{-\imaginaryi m \Omega \Delta t}
            \ProjectorP
            +
            \napier^{\hat{B}_m \Delta t} \hat{B}_m \DrazinInverse{\hat{B}_m}
            \ProjectorQ
            \nonumber \\
            &=
            \napier^{-\imaginaryi m \Omega \Delta t}\ProjectorP
            +
            \napier^{\hat{B}_m \Delta t} \ProjectorQ,
    \end{align}
    where we used $\Markov^{\mathrm{eff,vV}} \ProjectorP = \ProjectorP \Markov^{\mathrm{eff,vV}} = \hat{0}$.
    Therefore,
    $\hat{L}'(\Delta t) = \hat{R}'(\Delta t)$ and $\hat{L}(0) = \hat{R}(0) = \hat{0}$, hence $\hat{L}(\Delta t) = \hat{R}(\Delta t)$ holds for any $\Delta t \geq 0$~\cite{Arnold1978-nc}.

\section{Parameter Derivatives of Matrix Exponentials} \label{app:parameter_derivative_matrix_exponential}
    In this appendix, we summarize the parameter-derivative formula for matrix exponentials used in \cref{eq:time_averaged_current_op,eq:partial_U_general}.
    Let $A(\chi)$ be a differentiable square matrix.
    Then
    \begin{align}
        \partial_{\chi} \napier^{A(\chi)(t_1-t_0)}
            =
            \int_{t_0}^{t_1} \mathrm{d}s\,
            \napier^{A(\chi)(t_1-s)}
            \partial_\chi A(\chi)
            \napier^{A(\chi)(s-t_0)} .
        \label{eq:Duhamel_formula}
    \end{align}
    This formula is often referred to as Duhamel's formula or the Wilcox formula for parameter derivatives of exponential operators~\cite{Wilcox1967-nr,Peterson1967-qn,Bauer2013-hi}.

    To prove \cref{eq:Duhamel_formula}, we define
    \begin{align}
        F(s)
        =
        \napier^{A(\chi)(t_1-s)}
        \partial_{\chi} \napier^{A(\chi)(s-t_0)} .
    \end{align}
    Using
    \begin{align}
        \partial_s \napier^{A(\chi)(s-t_0)}
            &= A(\chi) \napier^{A(\chi)(s-t_0)}, \\
        \partial_s \napier^{A(\chi)(t_1-s)}
            &= - A(\chi) \napier^{A(\chi)(t_1-s)},
    \end{align}
    we have
    \begin{align}
        \partial_s F(s)
        =
        \napier^{A(\chi)(t_1-s)}
        \partial_{\chi} A(\chi)
        \napier^{A(\chi)(s-t_0)} .
    \end{align}
    Integrating from \(s=t_0\) to \(s=t_1\), and using
    \(F(t_0)=0\) and
    \(F(t_1)=\partial_{\chi}\napier^{A(\chi)(t_1-t_0)}\), we obtain
    \cref{eq:Duhamel_formula}.

    The same argument also applies to an operator $\hat{U}(\chi;s,t)$ satisfying
    \begin{align}
        \partial_s \hat{U}(\chi;s,t) &= \hat{W}(\chi;s) \hat{U}(\chi;s,t), \\
        \partial_t \hat{U}(\chi;s,t) &= -\hat{U}(\chi;s,t)\hat{W}(\chi;t).
    \end{align}
    In this case, the corresponding parameter-derivative formula is
    \begin{align}
        \partial_{\chi}\hat{U}(\chi;t_1,t_0)
        =
        \int_{t_0}^{t_1} \mathrm{d}s\,
        \hat{U}(\chi;t_1,s)
            \partial_{\chi}\hat{W}(\chi;s)
        \hat{U}(\chi;s,t_0).
        \label{eq:partial_U_general_formula}
    \end{align}

\section{Drazin Pseudoinverse} \label{app:Drazin_pseudoinverse}
  \subsection{Definition}
    The Drazin pseudoinverse $\DrazinInverse{A}$ of a square matrix $A$ is defined as the unique matrix satisfying the following three conditions \cite{Drazin1958-on,Horn2012-kv,Ben-Israel2003-ej}
    \begin{align}
      A \DrazinInverse{A} =& \DrazinInverse{A} A, \\
      \DrazinInverse{A} A \DrazinInverse{A} =& \DrazinInverse{A}, \\
      A^{k+1} \DrazinInverse{A} =& A^k,
    \end{align}
    where $k$ is the index of $A$, defined as the smallest non-negative integer such that $\mathrm{rank}(A^{k+1}) = \mathrm{rank}(A^k)$.

  \subsection{Properties}
    In the present two-state case, we consider a matrix $A=\Markov^{\mathrm{eff}, \mathrm{vV}}(0)$, which has a zero eigenvalue, and another eigenvalue with negative real part (see \cref{app:Perron_Frobenius_Theorem}).
    Therefore, we can set the index $k$ of $\Markov^{\mathrm{eff}, \mathrm{vV}}(0) \equiv \Markov$ to be $1$.
    Then, the Drazin pseudoinverse $\DrazinInverse{\Markov}$ can be expressed as
    \begin{align}
    \Markov \DrazinInverse{\Markov} &= \DrazinInverse{\Markov} \Markov = I - \ProjectorP, \\
    \ProjectorP \Markov &= \Markov \ProjectorP = \hat{0}, \\
    \ProjectorP^{2} &= \ProjectorP,
    \end{align}
    where $\ProjectorP$ is the projection operator onto the zero-eigenvalue subspace of $\Markov$.
    $I-\ProjectorP$ is the projection onto the complementary (invertible) spectral subspace of $\Markov$.

\section{Perron-Frobenius Theorem} \label{app:Perron_Frobenius_Theorem}
  \subsection{Irreducible Matrix}
    Let $A$ be a non-negative square matrix.
    $A$ is called \textit{irreducible} if for every pair of indices $(i, j)$, there exists a positive integer $k$ such that the $(i, j)$-th element of $A^k$ is positive.

    In the present irreducible two-state setting, the generator $\Markov^{\mathrm{eff}, \mathrm{vV}}(0)$ is a transition-rate matrix of a Markov process.
    Therefore, $\exp(t \, \Markov^{\mathrm{eff}, \mathrm{vV}}(0))$ is a non-negative square matrix for any $t \geq 0$ and is strictly positive for any $t > 0$.
    Hence, for any $t>0$, $\exp(t \, \Markov^{\mathrm{eff}, \mathrm{vV}}(0))$ is primitive, and in particular irreducible.

  \subsection{Perron-Frobenius Theorem}
    Let $M_n$ denote the set of all real $n \times n$ matrices.
    If ${A \in M_n}$ is an irreducible non-negative square matrix, then the following statements hold \cite{Horn2012-kv}:
    \begin{itemize}
      \item The spectral radius $r(A)$ of $A$ is a positive real eigenvalue of $A$.
      \item The eigenvalue $r(A)$ is simple (i.e., nondegenerate).
      \item There exist a unique (up to a multiplicative constant) pair of left and right eigenvectors with strictly positive components.
      \item Every other eigenvalue $\lambda$ of $A$ satisfies $|\lambda| \le r(A)$.
            If $A$ is \textit{primitive} (i.e., $A^k$ is strictly positive for some integer $k$), then $|\lambda| < r(A)$.
    \end{itemize}

    A matrix $P$ is called \textit{column-stochastic} if $\sum_i P_{ij} = 1$ and $P_{ij} \geq 0$ for all $i, j$.
    If $P$ is column-stochastic and irreducible, then the spectral radius of $P$ is $r(P) = 1$, the eigenvalue $1$ is simple, and all other eigenvalues $\mu$ of $P$ satisfy $|\mu| < 1$.

    Especially, if $\Markov \in M_n$ is a transition-rate matrix, then $P(t) = \exp(t \, \Markov)$ is column-stochastic for any $t\geq0$ and irreducible for any $t > 0$.
    Moreover, $P(t)$ is strictly positive for any $t > 0$, hence primitive.
    Therefore, $\Markov$ has a simple eigenvalue $0$, and all other eigenvalues $\lambda$ have negative real parts $\mathrm{Re}(\lambda) < 0$.